\documentclass[10pt,journal]{IEEEtran}
\IEEEoverridecommandlockouts
\usepackage{cite}
\usepackage{enumitem}
\usepackage{amsmath,amssymb,amsfonts}
\usepackage{graphicx}
\usepackage{amsmath}
\usepackage{xcolor}
\usepackage[normalem]{ulem}
\usepackage{multirow}
\usepackage{amssymb}
\newcommand\myworries[1]{\textcolor{blue}{#1}}
\usepackage{amsthm}

\newtheorem*{theorem*}{Theorem}

\newtheorem*{lemma*}{Lemma}
\usepackage{tabularx}
\usepackage{url}

\usepackage{algorithm}% http://ctan.org/pkg/algorithm
\usepackage{algpseudocode}% http://ctan.org/pkg/algorithmicx
\makeatletter
\def\therule{\makebox[\algorithmicindent][l]{\hspace*{.5em}\vrule height .75\baselineskip depth .25\baselineskip}}%

\newtoks\therules% Contains rules
\therules={}% Start with empty token list
\def\appendto#1#2{\expandafter#1\expandafter{\the#1#2}}% Append to token list
\def\gobblefirst#1{% Remove (first) from token list
	#1\expandafter\expandafter\expandafter{\expandafter\@gobble\the#1}}%
\def\LState{\State\unskip\the\therules}% New line-state
\def\pushindent{\appendto\therules\therule}%
\def\popindent{\gobblefirst\therules}%
\def\printindent{\unskip\the\therules}%
\def\printandpush{\printindent\pushindent}%
\def\popandprint{\popindent\printindent}%
\algdef{SE}[WHILE]{While}{EndWhile}[1]
{\printandpush\algorithmicwhile\ #1\ \algorithmicdo}
{\popandprint\algorithmicend\ \algorithmicwhile}%
\algdef{SE}[FOR]{For}{EndFor}[1]
{\printandpush\algorithmicfor\ #1\ \algorithmicdo}
{\popandprint\algorithmicend\ \algorithmicfor}%
\algdef{S}[FOR]{ForAll}[1]
{\printindent\algorithmicforall\ #1\ \algorithmicdo}%
\algdef{SE}[LOOP]{Loop}{EndLoop}
{\printandpush\algorithmicloop}
{\popandprint\algorithmicend\ \algorithmicloop}%
\algdef{SE}[REPEAT]{Repeat}{Until}
{\printandpush\algorithmicrepeat}[1]
{\popandprint\algorithmicuntil\ #1}%
\algdef{SE}[IF]{If}{EndIf}[1]
{\printandpush\algorithmicif\ #1\ \algorithmicthen}
{\popandprint\algorithmicend\ \algorithmicif}%
\algdef{C}[IF]{IF}{ElsIf}[1]
{\popandprint\pushindent\algorithmicelse\ \algorithmicif\ #1\ \algorithmicthen}%
\algdef{Ce}[ELSE]{IF}{Else}{EndIf}
{\popandprint\pushindent\algorithmicelse}%
\algdef{SE}[PROCEDURE]{Procedure}{EndProcedure}[2]
{\printandpush\algorithmicprocedure\ \textproc{#1}\ifthenelse{\equal{#2}{}}{}{(#2)}}%
{\popandprint\algorithmicend\ \algorithmicprocedure}%
\algdef{SE}[FUNCTION]{Function}{EndFunction}[2]
{\printandpush\algorithmicfunction\ \textproc{#1}\ifthenelse{\equal{#2}{}}{}{(#2)}}%
{\popandprint\algorithmicend\ \algorithmicfunction}%

\usepackage{array}
\newcolumntype{P}[1]{>{\centering\arraybackslash}p{#1}}

\usepackage{enumitem}
\newlist{legal}{enumerate}{10}
\setlist[legal]{label*=\arabic*.}

\newtheorem*{definition*}{Definition}

\newtheorem*{condition*}{Condition}
\newtheorem*{proof*}{Proof}

\renewcommand\myworries[1]{}
\usepackage{mathtools}
\newsavebox\myboxA
\newsavebox\myboxB
\newlength\mylenA

\newcommand{\subparagraph}{}

\newcommand*\xoverline[2][0.75]{%
	\sbox{\myboxA}{$\m@th#2$}%
	\setbox\myboxB\null% Phantom box
	\ht\myboxB=\ht\myboxA%
	\dp\myboxB=\dp\myboxA%
	\wd\myboxB=#1\wd\myboxA% Scale phantom
	\sbox\myboxB{$\m@th\overline{\copy\myboxB}$}%  Overlined phantom
	\setlength\mylenA{\the\wd\myboxA}%   calc width diff
	\addtolength\mylenA{-\the\wd\myboxB}%
	\ifdim\wd\myboxB<\wd\myboxA%
	\rlap{\hskip 0.5\mylenA\usebox\myboxB}{\usebox\myboxA}%
	\else
	\hskip -0.5\mylenA\rlap{\usebox\myboxA}{\hskip 0.5\mylenA\usebox\myboxB}%
	\fi}
\makeatother

\usepackage{scalerel}[2014/03/10]
\usepackage{stackengine}

\newlength\myindent
\usepackage{epsfig}
\usepackage{tabularx}
\usepackage{supertabular,booktabs}
\usepackage{capt-of}

\usepackage{longtable}
\usepackage{siunitx}
\usepackage{booktabs}

\usepackage{subcaption}

\usepackage{algorithm}
\usepackage{algpseudocode}
\renewcommand{\algorithmicforall}{\textbf{for each}}

\algrenewcommand\algorithmicrequire{\textbf{Precondition:}}
\algrenewcommand\algorithmicensure{\textbf{Postcondition:}}

\usepackage{etoolbox}
\usepackage{textcomp}
\usepackage[normalem]{ulem}
\usepackage{mathtools}
\usepackage{float}
\usepackage{tikz}
\usetikzlibrary{shapes.geometric, arrows}
\tikzstyle{startstop} = [rectangle, rounded corners, minimum width=3cm, minimum height=1cm,text centered, draw=black, fill=red!30]
\tikzstyle{io} = [trapezium, trapezium left angle=70, trapezium right angle=110, minimum width=3cm, minimum height=1cm, text centered, text width=3cm, draw=black, fill=blue!30]
\tikzstyle{process} = [rectangle, minimum width=3cm, minimum height=1cm, text centered, draw=black, fill=orange!30]
\tikzstyle{decision} = [diamond, minimum width=2.5cm, minimum height=2.5cm, text centered, draw=black, fill=green!30]
\tikzstyle{arrow} = [thick,->,>=stealth]
\tikzstyle{process} = [rectangle, minimum width=3cm, minimum height=1cm, text centered, text width=3cm, draw=black, fill=orange!30]
\def\BibTeX{{\rm B\kern-.05em{\sc i\kern-.025em b}\kern-.08em
		T\kern-.1667em\lower.7ex\hbox{E}\kern-.125emX}}
		
\usepackage[compact]{titlesec}
\titlespacing{\section}{0pt}{*0}{*0}
\titlespacing{\subsection}{0pt}{*0}{*0}
\titlespacing{\subsubsection}{0pt}{*0}{*0}
	
\begin{document}
	\bstctlcite{IEEEexample:BSTcontrol}
	
	% Distributed WAM based voltage stability monitoring using powerflow circles
	%Analytically Exact Distributed Voltage Stability Monitoring Scheme Using Power Flow Circles
	%\title{A Robust Decentralized Voltage Stability Index
	\title{Local Measurement Based Robust Voltage Stability Index \& Identification of Voltage Collapse Onset%PMU-based Online Local Monitoring of Small and Large Disturbance Long-term Voltage Stability%Robust Local Monitoring Scheme for Real-time Long-term Voltage Stability and Identification of Voltage Emergency Situations%\\
		%\thanks{This work is supported by the National Science Foundation award 1810537, U.S Department of Energy's (DOE's) ASSIST award (2019-2022), and  DOE's Advanced Research Projects Agency-Energy OPEN award (2019-2022).} \thanks{K. P. Guddanti and Y. Weng are with the School of Electrical, Computer and Energy Engineering at Arizona State University, emails:\{kguddant,yang.weng\}@asu.edu; A. R. R. Matavalam is with Department of Electrical and Computer Engineering at Iowa State University, email: amar@iastate.edu.}
	}

\makeatletter	
\patchcmd{\@maketitle}
  {\addvspace{0.5\baselineskip}\egroup}
  {\addvspace{-1\baselineskip}\egroup}
  {}
  {}
%	Kishan Prudhvi Guddanti, Yang Weng, Amarsagar Reddy Ramapuram Matavalam,Student Member

	\author{
		\IEEEauthorblockN{Kishan Guddanti},
		\IEEEauthorblockA{\textit{Student Member}},
		\textit{IEEE},
		\and
		\IEEEauthorblockN{Amarsagar Matavalam},
		\IEEEauthorblockA{\textit{Member}},
		\textit{IEEE},
		\and
		\IEEEauthorblockN{Yang Weng},
		\IEEEauthorblockA{\textit{Member}},
		\textit{IEEE}
	   % \and 
	   % \\ [-10.0em] 
	}
	%\DeclarePairedDelimiter\abs{\lvert}{\rvert}
	%\makeatletter
	%\let\oldabs\abs
	%\def\abs{\@ifstar{\oldabs}{\oldabs*}}
	
	%\IEEEaftertitletext{\vspace{-1\baselineskip}}
	
	\maketitle
%--------------------------------------------------------------------------------------------------------------------------	
	\begin{abstract}
	% oltage stability is critical to power grids and if leftunnoticed  it  can  lead  to  loss  of  load,  islanding  or  even  a  gridoutage. 
This paper addresses the problem of real-time monitoring of long-term voltage instability (LTVI) by using local field measurements. Existing local measurement-based methods use Thevenin equivalent parameter estimation that is sensitive to the noise in measurements. For solving this issue, we avoid the Thevenin approach by projecting the power flow equations as circles to develop local static-voltage stability indicator (LS-VSI) that is robust to measurement noises. The proposed method is an attractive option for practical implementation because of its decentralized nature, robustness to noise, and realistic modeling. Next, we utilize LS-VSI to propose a new local dynamic - VSI (LD-VSI) to identify the LTVI triggered by large disturbances and load tap changer dynamics. LD-VSI can identify not only the onset of instability with an alarm but also the voltage collapse point (VCP). Extensive numerical validation comparing LS-VSI with existing methods is presented on IEEE $30$-bus and larger test systems like $2000$-bus Texas synthetic grid to validate the robustness, accuracy, and situational awareness feature. We also verify the LD-VSI behavior on the Nordic power grid using PSSE dynamic simulation. When compared to existing methods, we observe that LD-VSI is not only more robust to measurement noise but also can identify VCP. 

	\end{abstract}
	
% 	\begin{IEEEkeywords}
% 		 voltage stability, wide-area monitoring and control, decentralized, Phasor measurement units, voltage collapse. 
% 	\end{IEEEkeywords}
% 	\vspace{-4mm}
	\section{introduction}
	\label{sec:intro}
%	\input{Intro.tex}
	%The advent of phasor measurement units (PMUs) and their wide-spread acceptance has made it possible to obtain real-time and time-synchronized information of voltage and current phasors for use in diversified applications such as monitoring and coordinated control actions in real-time for wide-area monitoring, protection, and control (WAMPAC) \cite{Phadke93,cui2019enhance}. 
%Among wide-area monitoring, protection, and control applications, online monitoring of long-term voltage instability (LTVI) is an area of interest for industry \cite{Novosel08}. Specifically, 
Long-term voltage instability (LTVI) is a quasi-static bifurcation (nose point of the PV curve) \cite{Cutsem98}, %caused by the inability of the generation and transmission system to provide sufficient power to loads, e.g., due to increased demand, generation outage, or generators on VAR limits \cite{Cutsem98}. 
if left unattended, LTVI results in a system-wide voltage collapse and blackout \cite{Cutsem98}. %,bollen03,taylor00}. 
Using field measurements, one can compute the margin (a measure of distance) from the current operating condition to LTVI. This measure of distance is quantified using metrics known as voltage collapse proximity indices (VCPIs). %This falls under wide-area monitoring, protection, and control applications \cite{Phadke93}. 
Based on the telecommunication requirement between the field measurement devices and control center, online monitoring of LTVI can be classified majorly into (a) wide-area VCPIs \cite{Wang11, Weng} and (b) local VCPIs \cite{Gubina95,Vu99,Milosevic03,Verbic04, Guddanti_tsg_ref}. Both approaches, wide-area and local, have a set of advantages and disadvantages which are well documented in \cite{Glavic11}. The online monitoring tools aim to aid the emergency control actions using system protection schemes (SPSs). %SPSs detect abnormal system conditions and take predetermined, corrective actions (other than the isolation of faulted elements) to preserve as far as possible system integrity and regain acceptable system performances \cite{van2007emergency}. 
Different online monitoring tools provide different types of output information that are used to trigger the SPSs. Based on the monitoring system type used \cite{van2007emergency}, SPSs are classified as wide-area SPSs (uses the output of wide-area VCPIs to trigger SPS) and local SPSs (uses the output of local VCPIs to trigger SPS). %\cite{van2007emergency} shows that local SPSs are more reliable as they do not rely on an extensive telecommunication system. Contrarily, local SPSs may lack the system-level perspective required to coordinate various, competing controls. 

A local SPS uses the output information from a local monitoring-type VCPI \cite{Vu99, Gubina95, Verbic04}. Local VCPIs calculate the two bus Thevenin equivalent \cite{Vu99, Gubina95, Verbic04} using real-time measurements at the bus of interest assuming \textit{constant power loads} \cite{barbier80, Milosevic03}. However, \cite{Milosevic03,pal1992voltage, concordia1982load, overbye1994effects} showed that voltage stability is impacted by the load characteristics (ZIP load models), and the assumption of constant power loads is not accurate depending on the mechanism of the instability event. Hence, the local VCPIs face two issues 1) the estimates of local VCPIs for LTVI margin, derived for a two-bus system assuming a constant power load, maybe pessimistic and inaccurate for the systems with mixed load types; 2) local VCPIs are highly impacted by the noise in measurements \cite{Guddanti_tsg_ref}.

% and the estimation of Thevenin equivalent parameters are highly impacted by the noise in the PMU measurements \cite{Guddanti_tsg_ref}.

% With regard to the constant power load assumption, unfortunately, voltage stability is affected by the load characteristics (static and dynamic) \cite{barbier80, pal1992voltage, concordia1982load, overbye1994effects}. 

%For example, when there are only constant power type loads in the system then the voltage-stability limit corresponds to the maximum power transfer limit (nose point of the PV curve). However, when voltage-sensitive loads are present in the system then the voltage stability limit may not occur at the maximum power limit anymore but rather beyond the maximum power limit (i.e., on the lower portion of the PV curve). Hence, the local VCPIs face two issues 1) the estimates of local VCPIs for LTVI margin, derived for a two-bus system assuming a constant power load, maybe pessimistic and inaccurate for the systems with mixed load types; 2) local VCPIs are highly impacted by the noise in the measurements \cite{Guddanti_tsg_ref}.

\underline{LS-VSI for small disturbance LTVI}: In this work, we propose a new local VCPI known as ``Local Static - Voltage Stability Indicator" (LS-VSI) that addresses the issues of 1) voltage-sensitive loads and 2) Thevenin equivalent parameters estimation (in the presence of noisy field measurements) by 1) incorporating the ZIP load dynamics into the LS-VSI formulation and 2) discarding the idea of obtaining two-bus Thevenin equivalent at the bus of interest, respectively.

% \textbf{\underline{Contributions of LS-VSI for small disturbance LTVI}}: In this work, we propose a new local VCPI known as ``Local Static - Voltage Stability Indicator" (LS-VSI) that addresses these issues as follows: \begin{enumerate}
%     \item LS-VSI incorporates the voltage-sensitive loads without constant power assumption like in \cite{barbier80, Milosevic03, Porco16}.
%     \item LS-VSI avoids Thevenin methodology while still retaining the single PMU requirement for local monitoring. Thus making it more robust to PMU noise.
% \end{enumerate}

Additionally, there is a different class of research focused on large disturbance long-term voltage stability phenomenon \cite{2004definition,2020definition}. This phenomenon considers the impact and interactions of nonlinear response due to devices such as motors, LTCs, generator field current limiters (OELs), and load restorative mechanisms (dynamic simulations). In case of a large disturbance phenomenon, \cite{vournas2008local} designed a local VCPI to output information (formula combining tap ratios of LTCs and voltages) which can be used to trigger SPSs. \cite{vournas2008local} showed that even though a voltage level-based VCPI is often a good indicator but it is not sufficient to provide a reliable, preventive picture of system security margins \cite{vournas2008local}. However, the methodology in \cite{vournas2008local} could not be extended to the buses without LTCs. \cite{vournas2016voltage} addressed the shortcomings of \cite{vournas2008local} by replacing the concept of LTCs with load conductance but its output is heavily impacted by the noise in measurement data as it uses a ``filter" (moving average over large time windows) as a preprocessor. For example, selecting a good threshold for the length of the filter's time window is a challenge. \textit{It can either trigger false alarms or delay the identification of the event or completely miss the occurrence of an event that recovered in a short period}.

{\underline{LD-VSI for large disturbance LTVI}}: In this work, we propose a second local-type VCPI for dynamic events arising from large disturbances. This proposed VCPI is known as ``Local dynamic - VSI (LD-VSI)" and it can be used to trigger an alarm indicating the onset of a voltage emergency situation. We addressed the shortcomings of \cite{vournas2008local} and \cite{vournas2016voltage}, by 1) designing the LD-VSI to be adaptable to any bus of interest and by 2) making LD-VSI less sensitive to the choice of the filter's window size, respectively. %Additionally, unlike in \cite{vournas2008local, vournas2016voltage}, the LD-VSI can also be used to identify the voltage collapse point (VCP). For example, the LD-VSI will tend to negative infinity at VCP. 

\begin{figure}
\begin{minipage}[t]{0.45\linewidth}
    \includegraphics[scale=0.3]{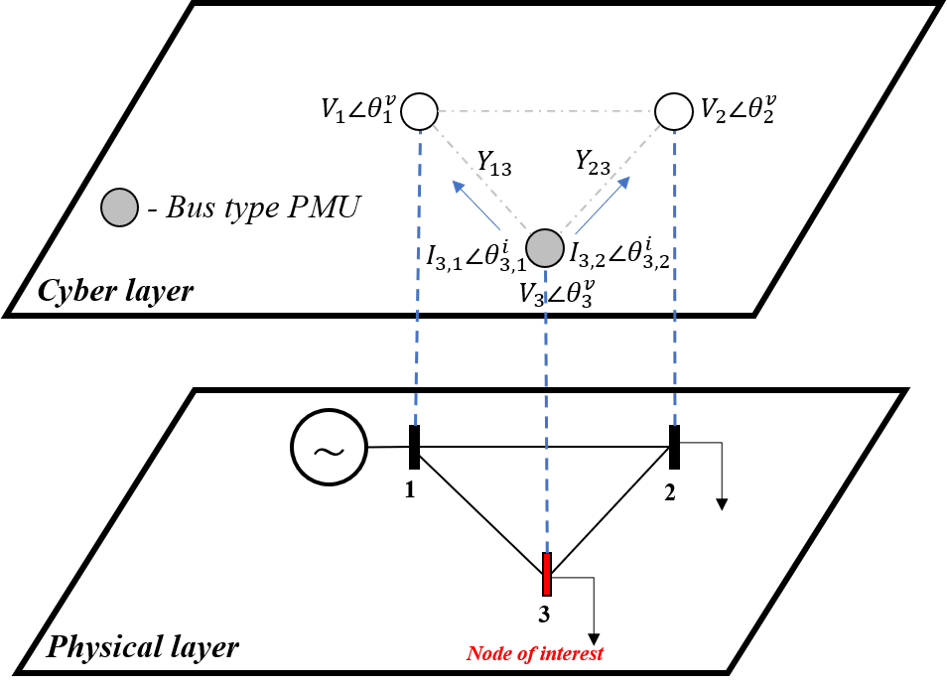}
    \caption{Decentralized/local monitoring architecture of grid.}
    \label{fig:decentralized_pic}
    \vspace{-1em}
\end{minipage}%
    \hfill%
\begin{minipage}[t]{0.45\linewidth}
    \hspace{1em}
    \includegraphics[scale=0.3]{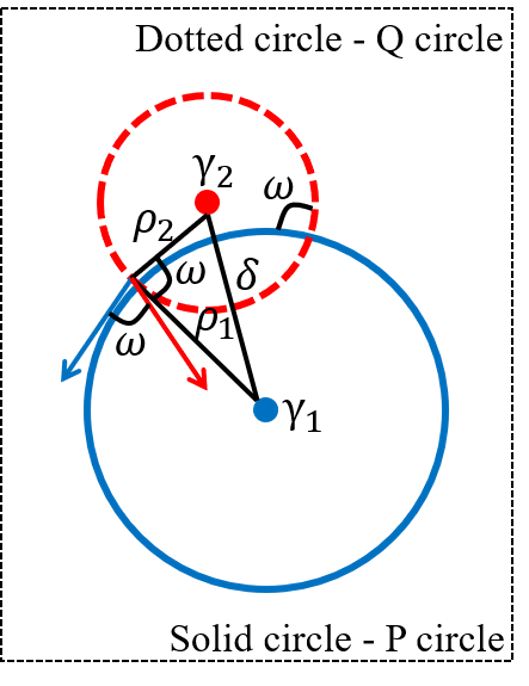}
    \caption{Power flow circles \eqref{eqn:pf_eqs_with_ltcs_shunts_zip} at bus $d$.}
    \label{fig:cosine_rule}
    \vspace{-1em}
\end{minipage}
\vspace{-1em}
\end{figure}

Both LS-VSI and LD-VSI are local measurement-based methods. For example, to monitor the LTVI phenomenon at bus $3$ for the system in  Fig.~\ref{fig:decentralized_pic}, using the LS-VSI and LD-VSI, we only need one measurement device at local bus $3$ to collect the bus $3$'s voltage phasor and adjacent lines' current phasor measurements (local measurements).

%Finally, we compare the proposed work with the existing methods in the literature to demonstrate its advantages. The LS-VSI is compared with a local VCPI \cite{Vu99} and a wide-area VCPI \cite{Wang11, Liu14} on IEEE $30$-bus system, highlighting LS-VSI advantages in 1) robustness to noisy PMU measurements (versus \cite{Vu99}) and 2) comparable performance (in terms of robustness to noisy PMU measurements) as that of a wide-area VCPI \cite{Wang11, Liu14} even though LS-VSI is a local-type VCPI. We also showed that LS-VSI can correctly identify the voltage collapse limit when the system has mixed load types (ZIP loads). \textit{Similar results are observed in the case of larger IEEE test systems as well (not included in the initial submission in the interest of space)}. Similarly, to demonstrate the advantage of LD-VSI, we used the IEEE standard test system developed by the voltage stability task force \cite{taskforce} for conducting dynamic long-term voltage stability studies i.e., IEEE Nordic system. We showed that the proposed LD-VSI is more robust to noise and has more flexibility over the selection of its filter's time window when compared with the New LIVES index (NLI) from \cite{vournas2016voltage}.

\textbf{\underline{Contributions of LS-VSI and LD-VSI}}: \begin{enumerate}
    \item LS-VSI incorporates the voltage-sensitive loads without constant power assumption like in \cite{barbier80, Milosevic03, Porco16}.
    \item LS-VSI is more robust to measurement noise than other Thevenin-based \cite{Vu99} local monitoring schemes.
    \item LD-VSI can be implemented at any load bus unlike the method in \cite{vournas2008local}.
    \item LD-VSI is less sensitive to the choice of filter's window size unlike NLI from \cite{vournas2016voltage}. Thus avoids false alarms and delayed event identification.
    \item Unlike the methods in \cite{vournas2008local, vournas2016voltage}, LD-VSI can identify (tends to minus infinity) the voltage collapse point (VCP).
    \item NLI \cite{vournas2016voltage} cannot be computed when the load conductance (G) decreases as it discards such data. LD-VSI does not have such an assumption and thus, it can be used to continuously monitor the grid regardless of G value. 
\end{enumerate}

The paper is organized as follows: Section~\ref{sec:impact_of_zip_loads} demonstrates the impact of voltage-sensitive loads on voltage stability limit. Section~\ref{sec:math_decentralized} derives the local static - VSI (LS-VSI) considering the ZIP load models, LTCs and presents the algorithm that uses LS-VSI to study LTVI. Section~\ref{sec:sim_1} presents the numerical validation and comparison of LS-VSI with other methods in the literature. Section~\ref{sec:nordic_system_ltcs} derives the local dynamic - VSI (LD-VSI) and presents the algorithm to use LD-VSI for LTVI studies. Section~\ref{sec:sim_2} presents the numerical validation and comparison of LD-VSI with other methods in the literature. Section~\ref{sec:conclusion} concludes the paper.

	\section{impact of voltage-sensitive loads on voltage stability limit}\label{sec:impact_of_zip_loads}
\begin{figure}
    \centering
    \includegraphics[scale=0.4]{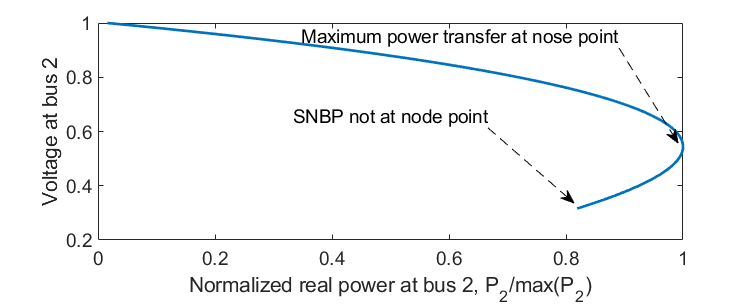}
    \caption{PV curve at bus $2$ when the load model is ZIP.}
    \label{fig:pv_curve_zip}
    \vspace{-2em}
\end{figure}

	It is known that the saddle-node bifurcation point (SNBP i.e., voltage collapse point) occurs at the nose point (maximum power transfer point) of the PV curve. However, this is not valid in the case of a power system with ZIP loads \cite{Cutsem98}. In this section, with the help of three bus example, as shown in Fig.~\ref{fig:pv_curve_zip}, we show that the SNBP may not always occur at the nose point where maximum power transfer occurs. 

\begin{figure}
    \centering
    \includegraphics[scale=0.4]{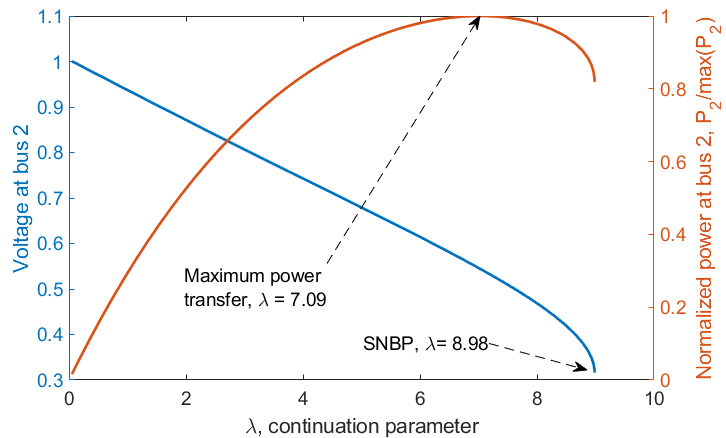}
    \caption{Load scaling parameter ($\lambda$, continuation parameter) versus voltage at bus $2$ when the load model is ZIP.}
    \label{fig:lmbda_v_curve_zip}
    \vspace{-1.7em}
\end{figure}

Let us consider a three bus fully connected mesh network with a slack bus at bus $1$, two ZIP loads at buses $2$ and $3$. Branch impedance of lines connecting buses $1$ to $3$ and $2$ to $3$ is $0.0074+j\cdot 0.0372$ p.u. and the impedance of line connecting buses $1$ to $2$ is $0.01+j\cdot 0.05$ p.u. The loads located at buses $2$ and $3$ are represented by ZIP load model whose Z, I, and P coefficients are $\alpha_p=0.8$, $\beta_p=0$, and $\gamma_p=0.2$ respectively. The real power at bus $2$ ($P_2$) is given by 
\begin{align}
    P_2 &= P_b\cdot \lambda \left[\alpha_p\cdot \|\overline{V}_2\|^2+\beta_p\cdot \|\overline{V}_2\|+\gamma_p\right], \label{eq:p2}
\end{align}
where $P_b$, $\overline{V}_2$, and $\lambda$ are the nominal load power, voltage phasor at bus $2$, and load scaling factor (continuation parameter in CPF-MATPOWER \cite{ZimmermanEtAl2011}) respectively. As shown in Fig.~\ref{fig:lmbda_v_curve_zip}, continuation power flow from MATPOWER is used to estimate the voltage at bus $2$ while increasing the load demand ($P_b\cdot \lambda$) at buses $2$ and $3$ by increasing the $\lambda$. From Fig.~\ref{fig:lmbda_v_curve_zip}, it can be observed that when $\lambda > 8.98$, the three bus system no longer has a power flow solution indicating the SNBP. 

In addition, it can be observed from Fig.~\ref{fig:lmbda_v_curve_zip} that the power at bus $2$ ($P_2$) is maximum when the load scaling factor $(\lambda)$ is $7.09$ but it is not maximum at the SNBP ($\lambda=8.98$). This is because from \eqref{eq:p2}, when load scaling factor ($\lambda$) increases the power injection at bus $2$ ($P_2$) increases. However, as the loading ($P_b \cdot \lambda$) on the grid increases the voltage $\overline{V}_2$ decreases. At sufficiently large loading on the grid (in this example when $\lambda=7.09$), the decrease in the magnitude of $\overline{V}_2$ (impedance nature of the load) dominates and reduces the net power injection at bus 2 ($P_2$). Fig.~\ref{fig:pv_curve_zip} shows the PV-curve demonstrating this phenomenon of SNBP not occurring at the maximum power transfer point (nose point of PV curve) when the power grid has ZIP loads. Additionally, as described in the latest $2020$ task force report on test systems for voltage stability \cite{taskforce}, \textit{other important parameters of interest that can onset a voltage instability phenomenon are the combined operation of load tap changers (LTCs) and over excitation limiters (OELs)}. \cite{vournas2008local, taskforce} demonstrates this phenomenon in a dynamic simulation environment to incorporate device (LTCs, OELs, etc.) induced stability mechanisms. \textit{\textbf{Hence it is important to consider the impact of not only ZIP load models but also the LTCs device mechanics in online measurement-based monitoring applications of voltage stability.}}%The recent literature \cite{Porco16,Guddanti_tsg_ref} proposed local-type PMU-based VCPIs to estimate the margin to voltage collapse. \textit{{However, they assume that the loads are $100\%$ constant power type loads. When the grid under investigation constitutes a portion of mixed type loads (ZIP loads), this assumption incorrectly equates the nose point (maximum power point) of the PV curve to be the SNBP.}}

% Unfortunately, the monitoring methods such as \cite{vournas2008local,vournas2016voltage} that consider this modeling are not robust to noisy PMU measurements, and the recent robust monitoring methods like \cite{Guddanti_tsg_ref, Porco16} do not consider the impact of LTCs and ZIP loads.}}

% When the system has ZIP loads and the load demands are not increasing then an improper operation of LTCs can result in a voltage collapse situation where the SNBP does not occur at the nose point of the PV curve . Thus, it is important to consider load models such as ZIP loads and LTCs in real-time monitoring applications of voltage stability.

% In addition to addressing the above shortcomings of \cite{Guddanti_tsg_ref,Porco16}, this paper proposes a robust decentralized/local monitoring method requiring only one PMU. Two monitoring algorithms are presented, one for the static power system without LTCs, restorative loads, etc. Another for a dynamic power system with LTCs that regulate the voltages.

% With this example as an introduction, we propose the decentralized VSI with the contributions highlighted in Section.~\ref{sec:intro}.

	%--
	\section{derivation of LS-VSI}
	\label{sec:math_decentralized}
	In this section, we derive a new VCPI that requires only local measurements (decentralized) obtained from a single measurement device (located at the bus of concern) known as a local static voltage stability indicator (LS-VSI). This indicator will consider critical parameters such as ZIP loads and LTCs while determining the distance to voltage collapse. First, we propose a new set of power flow equations that can incorporate critical parameters like the ZIP load model and load tap changers (LTCs). Second, we recast the derived power flow equations with detailed modeling of ZIP loads and LTCs into a decentralized/local framework. Finally, use these new equations to derive the proposed decentralized/local LS-VSI.

\subsection{Power Flow Equations with ZIP loads and LTCs}
\label{subsec:pf_as_circles}
Let $p_d$ and $q_d$ be the active and reactive power injections at bus $d$. Let  $\mathit{\overline{Y}}_{d,k} = \mathit{g}_{d,k} + j\cdot \mathit{b}_{d,k}$ be the $(d,k)^{th}$ element of the admittance matrix $Y$. {Let} $\mathit{v}_{d,r}$ and $\mathit{v}_{d,i}$ be the real and imaginary {parts} of the voltage phasor ($\overline{V}_d$) at bus $d$, respectively. We represent bus $d$'s neighboring bus set as $\mathcal{N}(\mathit{d})$. In the following subsections, LTCs and ZIP load model incorporated power flow equations in rectangular voltage coordinate are derived. \textit{In the interest of space, the derivation steps are not included in the initial submission}.%The power flow equations used in this paper are in rectangular coordinate form represented as circles \cite{Guddanti_tsg_ref}. As explained in \cite{Guddanti_tsg_ref}, this circle-based formulation helps to move away from Thevenin-based approach that can be sensitive to noisy measurements \cite{Vu99}. The power flow circle equations are given by
\subsubsection{Load Tap Changers and Shunts}
\label{subsubsec:ltc_shunts_addition}
\begin{figure}
    \centering
    \includegraphics[scale=0.425]{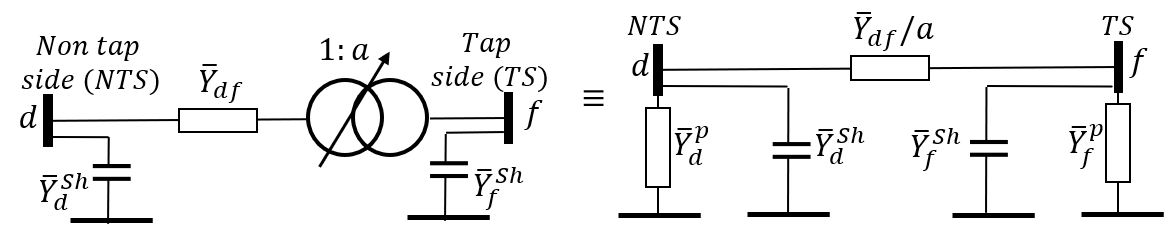}
    \caption{$\pi$ equivalent of load tap changing transformer.}
    \label{fig:ltc_pic}
    \vspace{-2em}
\end{figure}

As shown in Fig.~\ref{fig:ltc_pic}, let us consider an LTC connecting buses $d$ and $f$, buses $d$ and $f$ are on non-tap and tap side of the LTC respectively. $\overline{Y}_{df}$ be its short circuit admittance, $\overline{Y}^{Sh}_{d}=g^{Sh}_d+j\cdot b^{Sh}_d$ and $\overline{Y}^{Sh}_{f}=g^{Sh}_f+j\cdot b^{Sh}_f$ are the shunt admittances at buses $d$ and $f$ respectively. The $\pi$ equivalent representation of the LTC with shunts is presented in Fig.~\ref{fig:ltc_pic} on the right hand side. 

$\overline{Y}^p_{d}=g^{p}_d+j\cdot b^{p}_d=\overline{Y}_{df}\cdot (a-1)/a$, and $\overline{Y}^p_{f}=g^{p}_f+j\cdot b^{p}_f=\overline{Y}_{df}\cdot (1-a)/a^2$ represent the shunt equivalents of LTC on non-tap and tap side respectively. The neighboring bus set of bus $d$ is $\mathcal{N}(d)=\{f\}$. The conjugate operation is denoted by $()^*$. The net apparent power injection at bus $d$ is given by
% \begin{align*}
%     \overline{S}_d &= \overline{V}_d\cdot\left(\sum_{\mathit{k}\in\mathcal{N}(d)} \overline{Y}_{d,k}\cdot \overline{V}_{dk} + \overline{Y}^p_{d}\cdot \overline{V}_d + \overline{Y}^{Sh}_{d}\cdot \overline{V}_d\right)^*\\
%     &=\overline{V}_d\cdot \left(\sum_{\mathit{k}\in\mathcal{N}(d)} \overline{Y}_{d,k}\cdot \overline{V}_{dk}\right)^* + \overline{V}_d\cdot \left(\overline{Y}^p_{d}\cdot \overline{V}_d\right)^* + \overline{V}_d\cdot \left(\overline{Y}^{Sh}_{d}\cdot \overline{V}_d\right)^* \\
%     &= \overline{V}_d\cdot \left(\sum_{\mathit{k}\in\mathcal{N}(d)} \overline{Y}_{d,k}\cdot \overline{V}_{dk}\right)^* + \overline{V}_d\cdot \overline{V}^*_d\cdot \left(\overline{Y}^{p*}_{d} + \overline{Y}^{Sh*}_{d}\right)\\
%     &= \overline{V}_d\cdot \left(\sum_{\mathit{k}\in\mathcal{N}(d)} \overline{Y}_{d,k}\cdot \overline{V}_{dk}\right)^* + \left(v^2_{d,r}+v^2_{d,i}\right)\cdot\left(g^p_d + g^{Sh}_d - j\cdot \left(b^p_d + b^{Sh}_d\right)\right)
% \end{align*}
\[\begin{array}{l}
{{\bar S}_d} = {{\bar V}_d} \cdot {\left( {\sum\limits_{k \in {\cal N}(d)} {{{\bar Y}_{d,k}}}  \cdot {{\bar V}_{dk}} + \bar Y_d^p \cdot {{\bar V}_d} + \bar Y_d^{Sh} \cdot {{\bar V}_d}} \right)^*},\\
%  = {{\bar V}_d} \cdot {\left( {\sum\limits_{k \in {\cal N}(d)} {{{\bar Y}_{d,k}}}  \cdot {{\bar V}_{dk}}} \right)^*} + {{\bar V}_d} \cdot {\left( {\bar Y_d^p \cdot {{\bar V}_d}} \right)^*} + \\ {{\bar V}_d} \cdot {\left( {\bar Y_d^{Sh} \cdot {{\bar V}_d}} \right)^*},\\
 = {{\bar V}_d} \cdot {\left( {\sum\limits_{k \in {\cal N}(d)} {{{\bar Y}_{d,k}}}  \cdot {{\bar V}_{dk}}} \right)^*} + {{\bar V}_d} \cdot \bar V_d^* \cdot \left( {\bar Y_d^{p*} + \bar Y_d^{Sh*}} \right),\\
 = {{\bar V}_d} \cdot {\left( {\sum\limits_{k \in {\cal N}(d)} {{{\bar Y}_{d,k}}}  \cdot {{\bar V}_{dk}}} \right)^*} + \,\left( {v_{d,r}^2 + v_{d,i}^2} \right) \cdot \\ \left(
g_d^p + g_d^{Sh} -j \cdot \left( {b_d^p + b_d^{Sh}} \right) \right).
\end{array}\]
Let $\overline{S}_d=p_d+j\cdot q_d$ and upon further simplification, we get the new power flow equations in rectangular voltage coordinates considering both the LTCs and shunts as shown below.
% % \begin{align}
% %     p_d &= \Re\left(\overline{V}_d\cdot \left(\sum_{\mathit{k}\in\mathcal{N}(d)} \overline{Y}_{d,k}\cdot \overline{V}_{dk}\right)^*\right) + \left( {v_{d,r}^2 + v_{d,i}^2} \right) \cdot \left(g_d^p + g_d^{Sh}\right),\\
% %     q_d &= \Im{\left(\overline{V}_d\cdot \left(\sum_{\mathit{k}\in\mathcal{N}(d)} \overline{Y}_{d,k}\cdot \overline{V}_{dk}\right)^*\right)} + \left( {v_{d,r}^2 + v_{d,i}^2} \right) \cdot \left( {b_d^p + b_d^{Sh}} \right),
% % \end{align}
% \begin{align}
% &{p_d} = \operatorname{\mathbb{R}e}\left\{ {{{\bar V}_d} \cdot {{\left( {\sum\limits_{k \in {\cal N}(d)} {{{\bar Y}_{d,k}}}  \cdot {{\bar V}_{dk}}} \right)}^*}} \right\} + \left( {v_{d,r}^2 + v_{d,i}^2} \right) \cdot \notag \\
% &\left( {g_d^p + g_d^{Sh}} \right),\label{eq:temp_1}\\
% &{q_d} = \operatorname{\mathbb{I}m}\left\{ {{{\bar V}_d} \cdot {{\left( {\sum\limits_{k \in {\cal N}(d)} {{{\bar Y}_{d,k}}}  \cdot {{\bar V}_{dk}}} \right)}^*}} \right\} + \left( {v_{d,r}^2 + v_{d,i}^2} \right) \cdot \notag \\
% &\left( {b_d^p + b_d^{Sh}} \right).\label{eq:temp_2}
% \end{align}
\begin{subequations}\label{eqn:pf_eqs_with_ltcs_shunts}
\begin{equation}
\mathit{p}_d
= w_{d,1} \cdot \mathit{v}^2_{d,r} + w_{d,2} \cdot \mathit{v}_{d,r} + w_{d,1} \cdot \mathit{v}^2_{d,i} + w_{d,3} \cdot \mathit{v}_{d,i},    
\label{math_p_ltc}
\end{equation}
\begin{equation}
\mathit{q}_d
= w_{d,4} \cdot \mathit{v}^2_{d,r } - w_{d,3} \cdot \mathit{v}_{d,r} + w_{d,4} \cdot \mathit{v}^2_{d,i} + w_{d,2} \cdot \mathit{v}_{d,i}.    
\label{math_q_ltc}
\end{equation}
\end{subequations}
The parameters $w_{d,1},\ w_{d,2},\ w_{d,3}$ and $w_{d,4}$, considering the LTCs and shunts are given by
\begin{subequations}\label{eqn:parameters_ltc}
\begin{align}
w_{d,1} 
&= - \sum_{\mathit{k}\in\mathcal{N}(\mathit{d})} \mathit{g}_{d,k} + g_d^p + g_d^{Sh}, w_{d,4} 
=  \sum_{\mathit{k}\in\mathcal{N}(\mathit{d})} \mathit{b}_{d,k} - b_d^p \notag \\ &- b_d^{Sh}
\label{ref1_ltc}\\ w_{d,2}
&= \sum_{\mathit{k}\in\mathcal{N}(\mathit{d})} (\mathit{v}_{k,r}\mathit{g}_{d,k} - \mathit{v}_{k,i}\mathit{b}_{d,k}), w_{d,3}
= \sum_{\mathit{k}\in\mathcal{N}(\mathit{d})} (\mathit{v}_{k,r}\mathit{b}_{d,k} \notag\\ &+ \mathit{v}_{k,i}\mathit{g}_{d,k})
 \label{ref2_ltc}.%\\
% w_{d,3}
% &= \sum_{\mathit{k}\in\mathcal{N}(\mathit{d})} (\mathit{v}_{k,r}\mathit{b}_{d,k} + \mathit{v}_{k,i}\mathit{g}_{d,k}) ,\label{ref3_ltc}
% \\ w_{d,4} 
% &=  \sum_{\mathit{k}\in\mathcal{N}(\mathit{d})} \mathit{b}_{d,k} - b_d^p - b_d^{Sh}.
% \label{ref4_ltc}      
\end{align}
\end{subequations}
%Even though the power flow equations in \eqref{eqn:pf_eqs_with_ltcs_shunts} are more detailed (includes LTCs) than the equations presented in \cite{Weng, Porco16, Guddanti_tsg_ref}. Similar to \cite{Weng, Porco16, Guddanti_tsg_ref}, \eqref{eqn:pf_eqs_with_ltcs_shunts} is not sufficient to identify the SNBP of a power system containing ZIP loads due to its constant power load model assumption. Hence, 
\eqref{eqn:pf_eqs_with_ltcs_shunts} is modified to include the ZIP load model as shown below. %Later in this paper, using this ZIP load model and LTC based power flow circles, a new decentralized LTVI monitoring scheme requiring only one PMU is proposed.
\subsubsection{ZIP Load Models}
\label{subsubsec:zip_load_addition}
The real and reactive power demand at a bus $d$ with ZIP load model is given by
\begin{subequations}\label{eq:zip_equations}
\begin{align}
    p_d &= p_{L}\cdot \left[\alpha_p\cdot \|\overline{V}\|^2+\beta_p\cdot \|\overline{V}\|+\gamma_p\right],\\
    q_d &= q_{L}\cdot \left[\alpha_q\cdot \|\overline{V}\|^2+\beta_q\cdot \|\overline{V}\|+\gamma_q\right],
\end{align}
\end{subequations}
where $p_L,\ q_L$ are the nominal real and reactive powers of the load respectively. $\overline{V}$ is the normalized voltage phasor. $p_d,\ q_d$ are the net real and reactive power injections at bus $d$ with a ZIP load. $\alpha_p,\ \beta_p,\ \gamma_p$ are the Z, I, P coefficients of real power respectively and $\alpha_q,\ \beta_q,\ \gamma_q$ are the Z, I, P coefficients of reactive power respectively. \eqref{eqn:pf_eqs_with_ltcs_shunts} is further modified to include ZIP load model as show below
\begin{align}
    &\mathit{p}_d = w_{d,1} \cdot \mathit{v}^2_{d,r} + w_{d,2} \cdot \mathit{v}_{d,r} + w_{d,1} \cdot \mathit{v}^2_{d,i} + w_{d,3} \cdot \mathit{v}_{d,i}, \notag \\
    &\implies p_{L}\cdot \left[\alpha_p\cdot \|\overline{V}\|^2+\beta_p\cdot \|\overline{V}\|+\gamma_p\right] = w_{d,1} \cdot \mathit{v}^2_{d,r} + \notag \\ &w_{d,2} \cdot \mathit{v}_{d,r} + w_{d,1} \cdot \mathit{v}^2_{d,i} + w_{d,3} \cdot \mathit{v}_{d,i},\notag \\
    &\implies p_{L}\cdot \left[\alpha_p\cdot \left(\mathit{v}^2_{d,r}+\mathit{v}^2_{d,i}\right)+\beta_p\cdot\left(\mathit{v}^2_{d,r}+\mathit{v}^2_{d,i}\right)^{0.5}+\gamma_p\right] = \notag \\ &w_{d,1} \cdot \mathit{v}^2_{d,r} + w_{d,2} \cdot \mathit{v}_{d,r} + w_{d,1} \cdot \mathit{v}^2_{d,i} + w_{d,3} \cdot \mathit{v}_{d,i}. \label{eq:arrange_1}
\end{align}
To obtain the circle representing formulation, the power flow equations must be in the form of a homogeneous quadratic equation. However, the real power equation \eqref{eq:arrange_1} does not appear as a circle by physical law when there is a constant current components $\beta_p, \beta_q\neq 0$. %Similarly, the reactive power equation does not appear as a circle by physical law when $\beta_q\neq 0$. 
We identify this condition ($\beta_p\neq 0$, and $\beta_q\neq 0$) as a degenerate condition where the power flow equations by physical law cannot be represented as circles anymore. %\textcolor{blue}{It is also observed that the constant current type loads are not common in power system analysis \textcolor{red}{cite} and hence the degenerate condition is not often encountered.} 
In this section, we propose a formulation for the ZIP load models whose constant current component is zero i.e., $\beta_p= 0$, and $\beta_q= 0$. \textit{In the interest of space, we omitted our solution for degenerate conditions in this initial submission.}

% However, later in Section~\ref{sec:degenerate_circles}, the degenerate condition (when constant current components are non-zero i.e., $\beta_p\neq 0$, and $\beta_q\neq 0$) is addressed by approximating the power flow equations as circles but with a \textcolor{blue}{little to no trade-off} in the accuracy to identify the SNBP.

When $\beta_p= 0$, and $\beta_q= 0$, the real power equation in \eqref{math_p_ltc_zip} is obtained by rearranging the terms in \eqref{eq:arrange_1}. %by taking $\mathit{v}^2_{d,r},\ \mathit{v}^2_{d,i},\ \mathit{v}_{d,r},\ \mathit{v}_{d,i}$ common. 
A similar approach is carried out to obtain the reactive power equation as well. They are as follows
\begin{subequations}\label{eqn:pf_eqs_with_ltcs_shunts_zip}
\begin{equation}
\mathit{p}_L\cdot \gamma_p
= h_{d,1} \cdot \mathit{v}^2_{d,r} + h_{d,2} \cdot \mathit{v}_{d,r} + h_{d,1} \cdot \mathit{v}^2_{d,i} + h_{d,3} \cdot \mathit{v}_{d,i},    
\label{math_p_ltc_zip}
\end{equation}
\begin{equation}
\mathit{q}_L\cdot \gamma_q
= h_{d,4} \cdot \mathit{v}^2_{d,r } - h_{d,3} \cdot \mathit{v}_{d,r} + h_{d,4} \cdot \mathit{v}^2_{d,i} + h_{d,2} \cdot \mathit{v}_{d,i}.    
\label{math_q_ltc_zip}
\end{equation}
\end{subequations}
The parameters $h_{d,1},\ h_{d,2},\ h_{d,3}$ and $h_{d,4}$, considering the ZIP load model, LTCs and shunts are given by
\begin{subequations}\label{eqn:parameters_ltc_zip}
\begin{align}
h_{d,1} 
&= - \sum_{\mathit{k}\in\mathcal{N}(\mathit{d})} \mathit{g}_{d,k} + g_d^p + g_d^{Sh} - p_L\cdot \alpha_p, h_{d,4} 
=  \sum_{\mathit{k}\in\mathcal{N}(\mathit{d})} \mathit{b}_{d,k} \notag \\ &- b_d^p - b_d^{Sh} - q_L\cdot \alpha_q
\label{ref1_ltc_zip}\\ h_{d,2}
&= \sum_{\mathit{k}\in\mathcal{N}(\mathit{d})} (\mathit{v}_{k,r}\mathit{g}_{d,k} - \mathit{v}_{k,i}\mathit{b}_{d,k}), h_{d,3}
= \sum_{\mathit{k}\in\mathcal{N}(\mathit{d})} (\mathit{v}_{k,r}\mathit{b}_{d,k} \notag\\ &+ \mathit{v}_{k,i}\mathit{g}_{d,k}).
\label{ref2_ltc_zip}%\\
% h_{d,3}
% &= \sum_{\mathit{k}\in\mathcal{N}(\mathit{d})} (\mathit{v}_{k,r}\mathit{b}_{d,k} + \mathit{v}_{k,i}\mathit{g}_{d,k}) ,\label{ref3_ltc_zip}\\ 
% h_{d,4} 
% &=  \sum_{\mathit{k}\in\mathcal{N}(\mathit{d})} \mathit{b}_{d,k} - b_d^p - b_d^{Sh} - q_L\cdot \alpha_q.
% \label{ref4_ltc_zip}      
\end{align}
\end{subequations}

Equation \eqref{eqn:pf_eqs_with_ltcs_shunts_zip} represents the final representation of power flow equations with ZIP load models and LTCs. Using this representation, we derive the proposed local VCPI for small disturbance analysis (LS-VSI) in the subsection below. % by 1) designing the local PMU-based monitoring scheme and 2) explaining the derivation process. 

% When compared to \cite{Weng,Porco16,Guddanti_tsg_ref}, the derived power flow equations \eqref{eqn:pf_eqs_with_ltcs_shunts_zip} include the ZIP load models and LTCs in its formulation without the assumption of constant power loads and no LTCs. \textbf{Therefore, the proposed index derived using \eqref{eqn:pf_eqs_with_ltcs_shunts_zip} can identify the SNBP of the power system with ZIP loads and LTCs when \cite{Weng,Porco16,Guddanti_tsg_ref} fails. The proposed index is derived in the subsection below, in contrast to \cite{Guddanti_tsg_ref}, we derived the LS-VSI by further reducing the required number of PMUs to a single PMU and also removed the communication requirement between PMUs while also ensuring that the index is robust to noise.}  

\subsection{Derivation of Local Static - Voltage Stability Indicator}
\label{subsec:decentralized_vsi_derivation}
In this section, we will first design the measurement-based local monitoring scheme using \eqref{eqn:pf_eqs_with_ltcs_shunts_zip}. We eliminate the neighboring bus voltages $\overline{V}_k = v_{k,r}+j\cdot v_{k,i}$ in \eqref{eqn:pf_eqs_with_ltcs_shunts_zip} by using the branch currents as shown in \eqref{eq:local_calculate}. %{For example, at a given snapshot we need neighboring bus voltages to compute the voltage solution at bus $d$. In real-world, this can achieved using PMU voltage measurements of neighboring buses communicating with their common bus $d$ \cite{Guddanti_tsg_ref}. However, we can remove the communication dependency by estimating neighboring bus voltages using the adjacent branch current measurements.}

\begin{equation}
    \overline{V}_k = \overline{V}_{d,m} - I_{d,k}\cdot Z_{d,k},\ \forall \mathit{k}\in\mathcal{N}(\mathit{d}), \label{eq:local_calculate}
\end{equation}
where $I_{d,k}$ is the branch current phasor measurement, $\overline{V}_{d,m}$ is the voltage phasor measurement at bus $d$ and $Z_{d,k}$ is the complex impedance of adjacent branch of bus $d$ connecting bus $k$. Substituting \eqref{eq:local_calculate} in \eqref{eqn:pf_eqs_with_ltcs_shunts_zip}, the proposed decentralized/local power flow equations are given by \eqref{eqn:pf_eqs_with_ltcs_shunts_zip} where its new updated parameters $h_{d,1},\ h_{d,2},\ h_{d,3},$ \& $h_{d,4}$ are shown below in \eqref{eqn:new_parameters}. 
\vspace{-1em}

% The proposed decentralized power flow equations with LTCs, shunts and ZIP load model are given by \eqref{eqn:pf_eqs_with_ltcs_shunts_zip} where its new updated parameters $h_{d,1},\ h_{d,2},\ h_{d,3},$ \& $h_{d,4}$ are presented below in \eqref{eqn:new_parameters}.
\begin{subequations}\label{eqn:new_parameters}
\begin{align}%\label{new_ref1}
&h_{d,1} 
=\sum_{\mathit{k}\in\mathcal{N}(\mathit{d})} -\mathit{g}_{d,k} - g_d^p - g_d^{Sh} + p_L\cdot \alpha_p, h_{d,4}
=  \sum_{\mathit{k}\in\mathcal{N}(\mathit{d})} \mathit{b}_{d,k} \notag \\ &- b_d^p - b_d^{Sh} - q_L\cdot \alpha_q\\
&h_{d,2}
= \begin{multlined}[t]
    \sum_{\mathit{k}\in\mathcal{N}(\mathit{d})} (\operatorname{\mathbb{R}e}\{\overline{V}_{d,m} - I_{d,k}\cdot Z_{d,k}\}\mathit{g}_{d,k} - \\
    \operatorname{\mathbb{I}m}\{\overline{V}_{d,m} - I_{d,k}\cdot Z_{d,k}\}\mathit{b}_{d,k}),
\end{multlined} \\
&h_{d,3}
= 
\begin{multlined}[t]
\sum_{\mathit{k}\in\mathcal{N}(\mathit{d})} (\operatorname{\mathbb{R}e}\{\overline{V}_{d,m} - I_{d,k}\cdot Z_{d,k}\}\mathit{b}_{d,k} + \\
\operatorname{\mathbb{I}m}\{\overline{V}_{d,m} - I_{d,k}\cdot Z_{d,k}\}\mathit{g}_{d,k}). 
\end{multlined} %\\
% &h_{d,4}
% =  \sum_{\mathit{k}\in\mathcal{N}(\mathit{d})} \mathit{b}_{d,k} - b_d^p - b_d^{Sh} - q_L\cdot \alpha_q.
\end{align}
\end{subequations}
\vspace{-1em}
% \begin{subequations}\label{eqn:new_pf_eqs}
% \begin{equation}
% \mathit{p}_d
% = h_{d,1} \cdot \mathit{v}^2_{d,r} + h_{d,2} \cdot \mathit{v}_{d,r} + h_{d,1} \cdot \mathit{v}^2_{d,i} + h_{d,3} \cdot \mathit{v}_{d,i},    
% \label{new_math_p}
% \end{equation}
% \begin{equation}
% \mathit{q}_d
% = h_{d,4} \cdot \mathit{v}^2_{d,r } - h_{d,3} \cdot \mathit{v}_{d,r} + h_{d,4} \cdot \mathit{v}^2_{d,i} + h_{d,2} \cdot \mathit{v}_{d,i}.    
% \label{new_math_q}
% \end{equation}
% \end{subequations}
% The new parameters $h_{d,1},h_{d,2},h_{d,3}$ and $h_{d,4}$ are given by

% \begin{figure}
%     \centering
%     \includegraphics[scale=0.5]{img_folder/existence_cond_pic.png}
%     \caption{Power flow circles \eqref{eqn:new_pf_eqs} at a local bus $d$ in ($v_{d,r},v_{d,i}$) coordinate space given the snapshot measurements i.e., $v_{d,m},I_{d,k}$ from the bus type PMU located at bus $d$.}
%     \label{fig:cosine_rule}
% \end{figure}
% \begin{align*}
%     &{{\gamma_1}}
%     = \left(\frac{-h_{d,2}}{2h_{d,1}}, \frac{-h_{d,3}}{2h_{d,1}}\right), {\gamma_2}
%     = \left(\frac{h_{d,3}}{2h_{d,4}}, \frac{-h_{d,2}}{2h_{d,4}}\right),\\
%     &\rho_1
%     = \sqrt{\frac{\mathit{p}_{d}}{h_{d,1}} +  \frac{{h_{d,2}}^2 + {h_{d,3}}^2}{4{h_{d,1}}^2}}, \rho_2
%     = \sqrt{\frac{\mathit{q}_d}{h_{d,4}} +  \frac{{h_{d,3}}^2 + {h_{d,2}}^2}{4{h_{d,4}}^2}}. 
% \end{align*}

We call the power flow equations in \eqref{eqn:pf_eqs_with_ltcs_shunts_zip} with new updated parameters ($h_{d,1},\ h_{d,2},\ h_{d,3},$ \& $h_{d,4}$) from \eqref{eqn:new_parameters} as decentralized/local because these transformed power flow equations are now in terms of measurements from a single measurement device located at the bus of interest $d$. A further explanation is given below:
\begin{enumerate}
    \item The measurement device located at bus $d$ can provide the measurements of voltage phasor measurement at bus $d$ ($\overline{V}_{d,m}$) and adjacent branches' current measurements connected to bus $d$ ($\overline{I}_{d,k}\forall\ k\in\mathcal{N}(\mathit{d})$) \textbf{as the system operating condition changes with time}.
    \item Using these measurements ($\overline{V}_{d,m}$, $\overline{I}_{d,k}$), the parameters ($h_{d,1},\ h_{d,2},\ h_{d,3},$ \& $h_{d,4}$) from \eqref{eqn:new_parameters} can be computed instantaneously and \textbf{tracked in real-time}.
    \item \textbf{Please note that we do not assume that the field measurements ($\overline{V}_{d,m}$, $\overline{I}_{d,k}$) are constant but rather when the system operating condition changes (different loading snapshots) then the field measurement device located at bus $d$ will automatically provide the real-world measurements reflecting the changes in the system operating conditions. \textit{The proposed method tracks the system state using the field measurements in real-time}. This is standard and followed by every measurement-based monitoring method in the literature \cite{Vu99, Verbic04, Milosevic03}.}
    \item Finally, \textbf{at a snapshot}, given the local field measurements ($\overline{V}_{d,m}$, $\overline{I}_{d,k}$) at bus $d$, the parameters ($h_{d,1},\ h_{d,2},\ h_{d,3},$ \& $h_{d,4}$) from \eqref{eqn:new_parameters} are calculated and hence, \eqref{eqn:pf_eqs_with_ltcs_shunts_zip} represents a homogeneous quadratic equations (circles) in the domain of $v_{d,r}$ and $v_{d,i}$. 
\end{enumerate}

%{Thus, the power flow equations in \eqref{eqn:pf_eqs_with_ltcs_shunts_zip} are now in terms of measurements from a local PMU located at the bus of interest $d$}. 
As described in point (4) above, at a given snapshot, power flow equations in \eqref{eqn:pf_eqs_with_ltcs_shunts_zip} can be visualized as power flow circles. %\textbf{\cite{Guddanti_tsg_ref} shows that the online monitoring of the distance to voltage collapse is achieved by calculating the distance between the power flow circles.} 
\textit{\cite{Guddanti_tsg_ref} shows that when the power system experiences voltage collapse then the distance between the power flow circles is zero (internally or externally touching circles)}. As shown in Fig.~\ref{fig:cosine_rule}, to derive the proposed LS-VSI, we use the law of cosines in the triangle formed by the centers of the power flow circles and their points of intersections to identify the distance to voltage collapse. From Fig.~\ref{fig:cosine_rule}, let the centers of the real and reactive power circles be $\gamma_1$ and $\gamma_2$ respectively. Let the radius of real and reactive power circles be $\rho_1$ and $\rho_2$ respectively. The distance between the centers of the power flow circles is given by $\delta$ i.e., $\|\gamma_1 - \gamma_2\|_2 = \delta$. The parameters ($\gamma_1$, $\gamma_2$, $\rho_1$, and $\rho_2$) are obtained from \eqref{eqn:new_parameters}. As shown in Fig.~\ref{fig:cosine_rule}, using the cosine rule, we get $\cos^2(\omega) = \dfrac{\beta_{12}^2}{\rho_1^2 \cdot \rho_2^2}$ where $\beta_{12} = \dfrac{\delta^2 - (\rho_1^2 + \rho_2^2)}{2}$, and $\omega$ is the angle between the radii of real and reactive power circles. \textbf{\textit{When the real and reactive power circles touch each other either externally or internally, it results in the voltage collapse i.e., SNBP, this background is presented in \cite{Guddanti_tsg_ref,Cutsem98}}}. In this paper, we use the property of $cos^2() \le 1$, with equality occurring at the SNBP. Upon further simplification, we obtain the proposed LS-VSI as shown in \eqref{eq:final_vsi}. The results of the proposed decentralized index can be interpreted by understanding its lower bound. When the power flow circles touch each other i.e., $\delta = \rho_1 \pm \rho_2$, the proposed LS-VSI becomes zero i.e., $\Pi_1 = 0$ indicating the SNBP. %Now, we propose local measurement-based voltage assessment tools using \eqref{eq:final_vsi}.

	\vspace{-2em}
\begin{align}
    &cos^2(\omega) \leq 1,\ \implies \rho_1^2 \cdot \rho_2^2 - \beta_{12}^2 \geq 0, \\
    &\begin{multlined}[t]
\implies \frac{{\left( {{h_{d,2}}^2 + {h_{d,3}}^2} \right)}}{{{8 \cdot h_{d,1}}^2}} + \frac{{\left( {{h_{d,2}}^2 + {h_{d,3}}^2} \right)}}{{8\cdot {h_{d,4}}^2}} + \frac{{0.5{\mkern 1mu} p}}{{{h_{d,1}}}} + \frac{{0.5{\mkern 1mu} q}}{{{h_{d,4}}}} + \\
\left( {\frac{{\left({{h_{d,2}}^2 + {h_{d,3}}^2} \right)}}{{4*{h_{d,1}}^2}} + \frac{p_L\cdot \gamma_p}{{{h_{d,1}}}}} \right)\cdot \left( {\frac{{\left( {{h_{d,2}}^2 + {h_{d,3}}^2} \right)}}{{4*{h_{d,4}}^2}} + \frac{q_L\cdot \gamma_q}{{{h_{d,4}}}}} \right) \\ - 0.5{\mkern 1mu} {\left| {\frac{{{h_{d,2}}}}{{{2{\mkern 1mu}h_{d,1}}}} + \frac{{{h_{d,3}}}}{{2{\mkern 1mu}{h_{d,4}}}}} \right|^2} - 0.5{\mkern 1mu} {\left| {\frac{{{h_{d,3}}}}{{2{\mkern 1mu}{h_{d,1}}}} - \frac{{{h_{d,2}}}}{{2{\mkern 1mu}{h_{d,4}}}}} \right|^2} \ge 0.\notag%~~~~~~~~~~~
\end{multlined}\\
    % \\ &\implies \Pi_1 \ge 0, \\
    & \therefore \text{Proposed LS-VSI} = \Pi_1 \ge 0.
    \label{eq:final_vsi}
    % \vspace{-2em}
\end{align}
\vspace{-2em}

Here, we propose an algorithm for local (decentralized) measurement-based long-term voltage stability monitoring. The algorithm to monitor the voltage stability of the power grid is presented in Algorithm~\ref{alg:algo}. \textbf{It is important to note that when \eqref{eq:final_vsi} is equal to zero, it represents the SNBP (voltage collapse) but not the nose point (maximum power point) of the PV curve as discussed in Section~\ref{sec:impact_of_zip_loads}.} Since step $1$ to step $4$ in Algorithm~\ref{alg:algo} are analytical equations, they can be computed nearly instantly with time complexity of $O(1)$, making it an attractive tool for online voltage stability assessment.% for small disturbance phenomena. Small disturbance phenomenon corresponds to small perturbations such as incremental changes in system load \cite{2004definition,2020definition}.

 %When the system has ZIP loads, we show in the simulation section that \cite{Guddanti_tsg_ref} fails due to its constant power assumptions while the proposed LS-VSI identifies the SNBP correctly. %\textbf{\textit{In this paper, the voltage/current phasor measurements ({$\overline{V}_d$}, {${I_{d,k}}$}) are not assumed to be constant for different loading conditions or snapshots. The proposed method is PMU measurement-based and it uses PMUs to continuously monitor the voltages and current in real-time similar to many other methods in the literature \cite{Vu99,Porco16,vournas2016voltage,Guddanti_tsg_ref}.}} 
 %(\textit{please note that as the system operating condition changes, the measurements from PMU reflect the change in system state in real-time and Algorithm~\ref{alg:algo} calculates the LS-VSI in real-time accordingly)}.

\vspace{-0.8em}
\begin{algorithm}[!h]
% \vspace{-2em}
\caption{Local monitoring of SNBP \textit{at a given snapshot}}\label{alg:algo}
\begin{algorithmic}[1]
  %  \Procedure{myalgo}{}
  \Statex \textbf{Input~~:} Measurements: $\overline{V}_d$, $I_{d,k}$, $Y_{d,k}$ $\forall\ k\ \in \mathcal{N}(d)$, $\alpha_p,\ \beta_p,\ \gamma_p$ and $\alpha_q,\ \beta_q,\ \gamma_q$ at bus $d$.
  \Statex \textbf{Output:} Proposed index (LS-VSI) at bus $d$.
  %\vspace{-1em}
  \State Calculate the net injection at bus $d$ i.e., $\overline{S}_d = p_d+j\cdot q_d$;
  \State Calculate the $p_L$ and $q_L$ using \eqref{eq:zip_equations};
  \State Calculate parameters $h_{d,1},h_{d,2},h_{d,3}$ and $h_{d,4}$ using \eqref{eqn:new_parameters};
  \State \Return Calculate the proposed LS-VSI ($\Pi_1$) using \eqref{eq:final_vsi}.
  %  \EndProcedure
\end{algorithmic}
\end{algorithm}

% \begin{algorithm}[!h]
% \caption{Distributed monitoring of SNBP \textit{at a given snapshot}}\label{alg:algo}
% \begin{algorithmic}[1]
%   %  \Procedure{myalgo}{}
%   \Statex \textbf{Input~~:} Measurements: $\overline{V}_{d,k}$, $I_{d,k}$, $Y_{d,k}$ $\forall\ k\ \in \mathcal{N}(d)$.
%   \Statex \textbf{Output:} Proposed index (VSI) at bus $d$.
%   %\vspace{-1em}
%   \State Calculate the voltage at bus $d$ i.e., $\overline{V}_d$.
%   \State Calculate the net injection at bus $d$ i.e., $\overline{S}_d = p_d+j\cdot q_d$;
%   \State Calculate parameters $t_{d,1},t_{d,2},t_{d,3}$ and $t_{d,4}$ using \eqref{eqn:new_parameters};
%   \State \Return Calculate the proposed VSI $\Delta_{norm}$.
%   %  \EndProcedure
% \end{algorithmic}
% \end{algorithm}
		%--
	\section{algorithm 2 using LD-VSI: local measurement based monitoring for large disturbance long-term voltage stability}
	\label{sec:nordic_system_ltcs}
	In Section~\ref{sec:math_decentralized}, we proposed a decentralized algorithm to monitor LTVI for small disturbance phenomenon and its advantageous features are discussed in Section~\ref{sec:sim_1}. In this section, we propose another local (decentralized) measurement based algorithm for long-term voltage stability monitoring but for a large disturbance phenomenon. Large disturbance phenomenon corresponds to impact and interaction of nonlinear response of devices such as transformer tap changers (LTCs), generator field current limiters (OELs), and load restorative mechanisms \cite{2004definition,2020definition}. We use the benchmark IEEE Nordic test system presented in the $2020$ Task Force report for voltage stability studies \cite{taskforce} to monitor the instability due to large disturbance phenomenon.
\subsection{Large disturbance long-term voltage instability phenomenon in IEEE Nordic test system}
\label{subsec:nordic_test_system}
\begin{figure}
    \centering
    % \vspace{-1em}
    \includegraphics[scale=0.35]{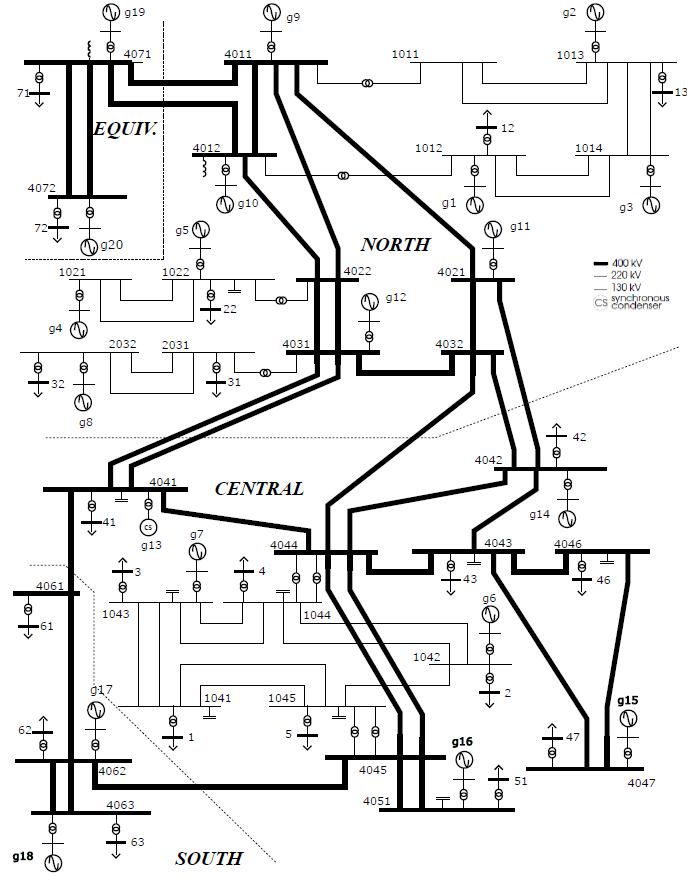}%[width=\linewidth]
    \caption{One-line diagram of Nordic test system.}
    \label{fig:nordic_pic}
    \vspace{-1em}
\end{figure}

\begin{figure}
    \centering
    % \vspace{-2em}
    \includegraphics[width = 0.65\linewidth]{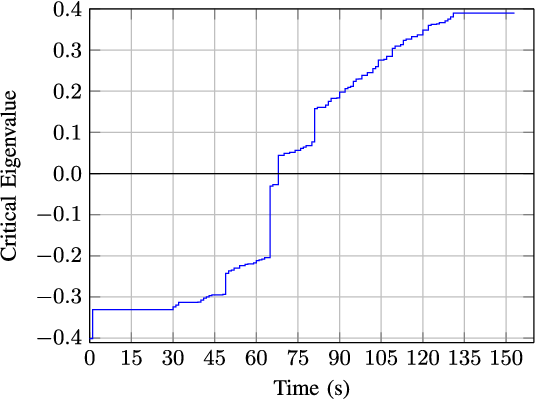}
    \caption{Dominant eigenvalue for unstable scenario \cite{vournas2016voltage}. Zero crossing at $63$s corresponds to onset of instability. }
    \label{fig:nordic_eigen}
    \vspace{-2em}
\end{figure}

Fig.~\ref{fig:nordic_pic} shows the IEEE Nordic test system. Detailed IEEE Nordic test system data, as well as operating points and responses
of this system to contingencies, are provided in \cite{taskforce} and all relevant files are downloadable from the PSDP Committee web page \cite{nordic_files}. Generators are modeled in detail including magnetic saturation, AVRs, and prime movers/governors. Active loads have both constant impedance ($\alpha_p=0.5$) and power ($\gamma_p=0.5$) components while the reactive loads have only constant impedance ($\alpha_q=1$) component. The IEEE Nordic test system has four area namely, 1) “North” with hydro generation and some load, 2) “Central” with much higher load and thermal power generation, 3) “Equiv” connected to the “North”, which includes a simple equivalent of an external system, and 4) “South” with thermal generation, which is rather loosely connected to the rest of the system. The IEEE Nordic test system is loaded in such a way that there are large transfers that flow from the North to Central areas. The Nordic test system experiences a long-term voltage instability phenomenon when a three-phase fault near bus $4032$ is cleared by tripping of the line $4032$-$4044$. The tripping of the line $4032$-$4044$ causes additional load demand to flow through the remaining tie lines resulting in a sudden distribution side voltage drop that is corrected by the LTCs \cite{taskforce}. However, as \underline{described in detail in \cite{taskforce}}, after the disturbance the maximum power that can be delivered by combined generation and transmission system is smaller than what the LTCs attempt to restore resulting in voltage instability phenomenon \cite{taskforce} (\textit{in the interest of space, the event is not fully explained in this initial draft but rather we refer to the work in \cite{taskforce} for more details about this event in Nordic power grid}). This instability phenomenon driven by OELs and LTCs takes place a few minutes after the initiating event is cleared as explained in-detail in \cite{taskforce}. %For example, as shown in Fig.~\ref{fig:LTC_vmag_pic}, after the disturbance is cleared, the LTC joining buses $1$ and $1041$ tries to restore the distribution side voltage at bus $1$ by requiring more reactive power support. Since the overexcitation limiters of the generators in the North reduce their reactive capabilities and the transfer capability is reduced due to the tie-line ($4032$-$4044$) trip, the attempts of the LTC fail to restore the voltage resulting in the voltage collapse. 
The plot of the dominant eigenvalue of the system versus time is shown in Fig. \ref{fig:nordic_eigen} demonstrating the onset of instability is around $63$s. In this paper, we propose a method to identify the onset of a voltage emergency situation.

% In this paper, we propose a second algorithm to identify the onset of a voltage emergency situation.% \cite{vournas2008local, vournas2016voltage} proposed methods to preemptively identify the onset of voltage collapse (SNBP) by triggering an alarm ahead of time. More specifically, 
%\cite{vournas2016voltage} proposed decentralized monitoring algorithms based on load conductance that identify the maximum power point which occurs before the SNBP. We propose an algorithm that has better features than \cite{vournas2016voltage} below.% to identify the onset of voltage collapse emergency situations in case of a large disturbance LTVI phenomenon that is triggered by the combined operation of LTCs and OELs \cite{taskforce}.
\begin{figure}
    \centering
    % \vspace{-1em}
    \includegraphics[width=\linewidth]{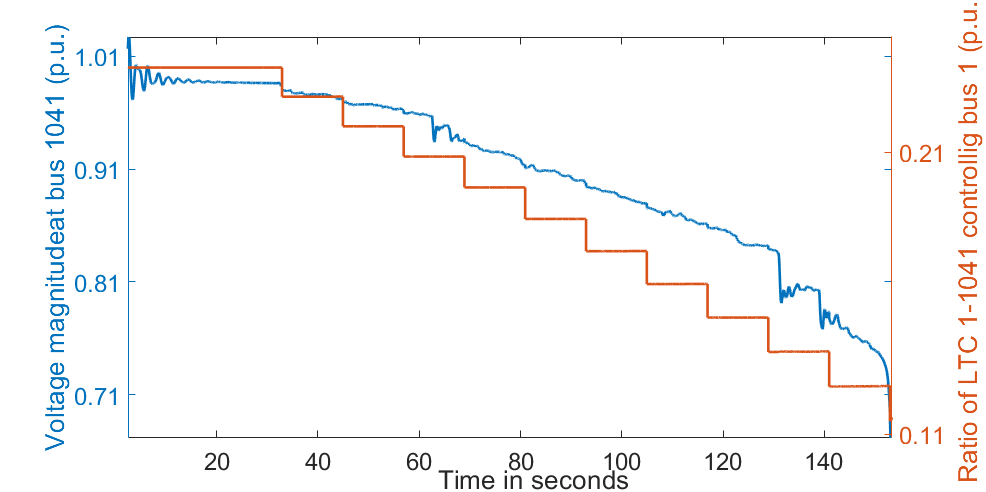}
    \caption{Failed restoration of distribution voltage by LTC.}
    \label{fig:LTC_vmag_pic}
    \vspace{-2em}
\end{figure}

\subsection{Characteristic curves to identify the maximum power point}
\label{subsec:charcateristic_curve}
In this subsection, first, we provide a simple three bus example to explain the working principle behind \cite{vournas2016voltage}. Second, using the same three bus example, we present the LD-VSI.

\begin{figure}
    \centering
    \vspace{-1em}
    \includegraphics[width=\linewidth]{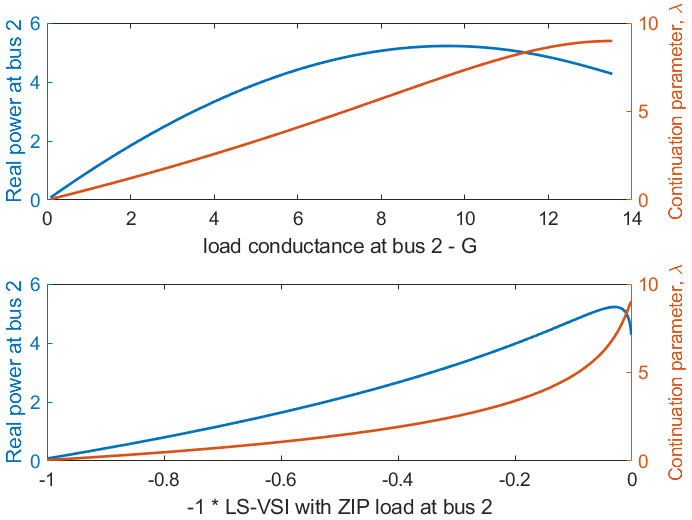}
    \caption{Characteristic curves to monitor the maximum power point (nose point). ($P_2$ versus $G_2$) for \cite{vournas2016voltage}, and ($P_2$ versus $-\Pi_1$) for proposed algorithm $2$.}
    \label{fig:charac_curves}
    \vspace{-1em}
\end{figure}

Let us consider the three bus fully connected network described in Section ~\ref{sec:impact_of_zip_loads}. The loads at buses $2$ ($P_2$) and $3$ ($P_3$) are increased by increasing the $\lambda$ from \eqref{eq:p2}. As shown in Fig.~\ref{fig:charac_curves}, \cite{vournas2016voltage} proposed New LIVES Index (NLI) identify the nose point (maximum power point) by monitoring the sign change of the slope of the real power demand ($P$) versus load conductance ($G$) in the characteristic curve. As shown in Fig.~\ref{fig:charac_curves}, we propose a new decentralized VCPI known as local dynamic - voltage stability indicator (LD-VSI) to identify the nose point (maximum power point) by monitoring the sign change of the slope of the real power demand ($P$) versus the negative sign of LS-VSI ($-\Pi_1$) characteristic curve. This approach works as the LS-VSI monotonically reduces as $\lambda$ increases. Thus, the change of sign of $\Pi_2$ from positive to negative corresponds to the maximum power point. 
In Fig.~\ref{fig:charac_curves}, it can be observed that both NLI \cite{vournas2016voltage} and the proposed index $\Pi_2$ identify the maximum power point at $\lambda=7.09$ when their sign changes from positive to negative. Thus, LD-VSI ($\Pi_2$) can be used to identify voltage collapse driven by a large disturbance phenomenon. The proposed decentralized index at bus $d$ to identify the maximum power point during a large disturbance LTVI phenomenon is given by

% The monitoring of the NLI=$\Delta P/\Delta G$ involves usage of a filter with window size as hyper parameter (equation $17$ in \cite{vournas2016voltage}). One . There are two main problems, 1) the window size hyper parameter plays very crucial role and an incorrect size results in false alarms which confuse the operator to trigger any protection schemes. \cite{taskforce}. 2) load conductance $G$ is calculated using the noisy voltage and current measurements (equation $6$ in \cite{vournas2016voltage}) which makes the NLI more sensitive to noise in PMU measurements, making it less robust.

\begin{equation}
    \text{LD-VSI} = \Pi_2|_\text{at bus $d$}  = \dfrac{\Delta\ P_d}{-\Delta\  \Pi_1|_{ \text{at bus $d$}}},%\ \dfrac{\Delta\ P_d}{\Delta\  G_d}% = \dfrac{\Delta\ P_d}{-\Delta\  \text{LS-VSI}|_{ \text{at bus $d$}}},
    \label{eq:final_vsi_2}
\end{equation}
where $\Delta$ indicates the filtering method described in \cite{vournas2016voltage}. The overall procedure is described in Algorithm \ref{alg:algo2}. %The difference between the NLI and proposed LD-VSI ($\Pi_2$) is in their denominators i.e., $\Delta G$ and $\Delta \Pi_1$ respectively.

\textbf{The key advantage of LD-VSI over NLI is that the slope transition is much more prominent for LD-VSI compared to NLI} due to the fact that its denominator (LS-VSI) is within a smaller range (0 to 1) compared to the load conductance G (0 to 13). \textbf{The large range of G makes the slope (NLI) smaller and sensitive to perturbations due to noise. This issue is reduced in LD-VSI ($\Pi_2$) making it more robust to noise compared to NLI.} A further advantage is that the characteristic curve P vs $\Pi_1$ becomes nearly vertical as the system approaches the SNBP (maximum $\lambda$). This is because the value of LS-VSI ($\Pi_1$) approaches zero at the critical bus near the SNBP. \textbf{Thus, LD-VSI ($\Pi_2$) can also be used to identify the SNBP of the system and the critical bus when the value of $\Pi_2$ approaches a large negative value. This cannot be done by using the LIVES and New LIVES indices in\cite{vournas2016voltage, vournas2008local}.} Furthermore, NLI can only be calculated when the load conductance increases at the bus of interest otherwise it discards the measurement data, this does not allow continuous monitoring of the grid. Whereas, the proposed LD-VSI does not have any such assumptions. Simulation discussion for LD-VSI is presented in Section~\ref{sec:sim_2}.

% \begin{remark}
% As demonstrated in Section.~\ref{sec:sim_1}, the VSI ($\Pi_1$) is highly robust to noise but $G= \operatorname{\mathbb{R}e}\left(\dfrac{\overline{I}}{\overline{V}}\right)$ is sensitive to noise due to the division of two noisy signals i.e., $\overline{I}$ and $\overline{V}$.
% \end{remark}

%Hence, the proposed decentralized index for large disturbance phenomenon $\Pi_2=\Delta P/ \Delta -\Pi_1$ is robust to noisy PMU data when compared to NLI=$\Delta P/\Delta G$ from \cite{vournas2016voltage}. \textbf{\textit{Therefore, the proposed decentralized index $\Pi_2$ requires smaller window sizes when compared to NLI from \cite{vournas2016voltage}, making the proposed index ($\Pi_2$) more favourable to use.}} The overall algorithm is presented below.

% \subsection{Decentralized algorithm 2 for large disturbance events}
% \label{subsec:algorithm2}
\vspace{-1em}
\begin{algorithm}
\caption{Detection of onset of voltage collapse for large disturbance events: local monitoring of maximum power point}\label{alg:algo2}
\begin{algorithmic}[1]
  %  \Procedure{myalgo}{}
  \Statex \textbf{Input~~:} Measurements: $\overline{V}_d$, $I_{d,k}$, $Y_{d,k}$ $\forall\ k\ \in \mathcal{N}(d)$, $\alpha_p,\ \beta_p,\ \gamma_p$ and $\alpha_q,\ \beta_q,\ \gamma_q$ at bus $d$, window length $T$.
  \Statex \textbf{Output:} Alarm to indicate onset of voltage collapse. 
  %\vspace{-1em}
  \State Calculate $\Pi_1$ for all time steps in time window $T$; %\Comment{Implement Algorithm~\ref{alg:algo} for all time steps in $T$.}
  \State Calculate $\Delta p_d$ and $\Delta \Pi_1|_\text{at bus $d$}$ using the filter in \cite{vournas2016voltage};
  \State Calculate LD-VSI ($\Pi_2$) at bus $d$ using \eqref{eq:final_vsi_2};
  \State \Return If $\Pi_2<0$ then trigger alarm.
  %  \EndProcedure
\end{algorithmic}
% \vspace{-2em}
\end{algorithm}
\vspace{-1em}
% \vspace{-2em}

% In this paper, when compared to the decentralized algorithm in \cite{vournas2016voltage}, we show that the proposed second algorithm in Section~\ref{sec:nordic_system_ltcs} requires smaller filter window size, and more robust to the noisy PMU measurements. Moreover, it is unclear how the 
	%--
	\section{simulation 1 in MATPOWER: small disturbance long-term voltage stability}
	\label{sec:sim_1}
	% \subsection{Proposed Decentralized VSI Versus Centralized, Other Decentralized VSIs}
% \label{subsec:table_comparison_subsection}
The proposed Algorithm~\ref{alg:algo} is tested on IEEE $30$-bus system \cite{matfile} and $2000$-bus Texas synthetic grid \cite{Birchfield17}. { Similar results are observed for IEEE $300$ and $2383$-bus (Polish) systems as well}. In the interest of space, the results for larger bus systems are not included. To obtain the voltage, current phasor measurements, and the true voltage stability margin, we used the non-divergent robust Newton-Raphson based power flow solver known as continuation power flow (CPF) from MATPOWER \cite{ZimmermanEtAl2011}. In this section, we show that the proposed decentralized LS-VSI in Algorithm~\ref{alg:algo} is advantageous compared to both centralized and decentralized VCPIs. The LS-VSI is shown to be highly robust to noisy measurements when compared with other decentralized and centralized VCPIs. We also show that LS-VSI \eqref{eq:final_vsi} can capture the impact of system disturbances that occur away from the monitored location. We also show that the proposed decentralized VSI can identify the critical bus SNBP accurately while \cite{Guddanti_tsg_ref} \cite{Porco16} fail.
\begin{figure}
    \centering
    \includegraphics[width=\linewidth]{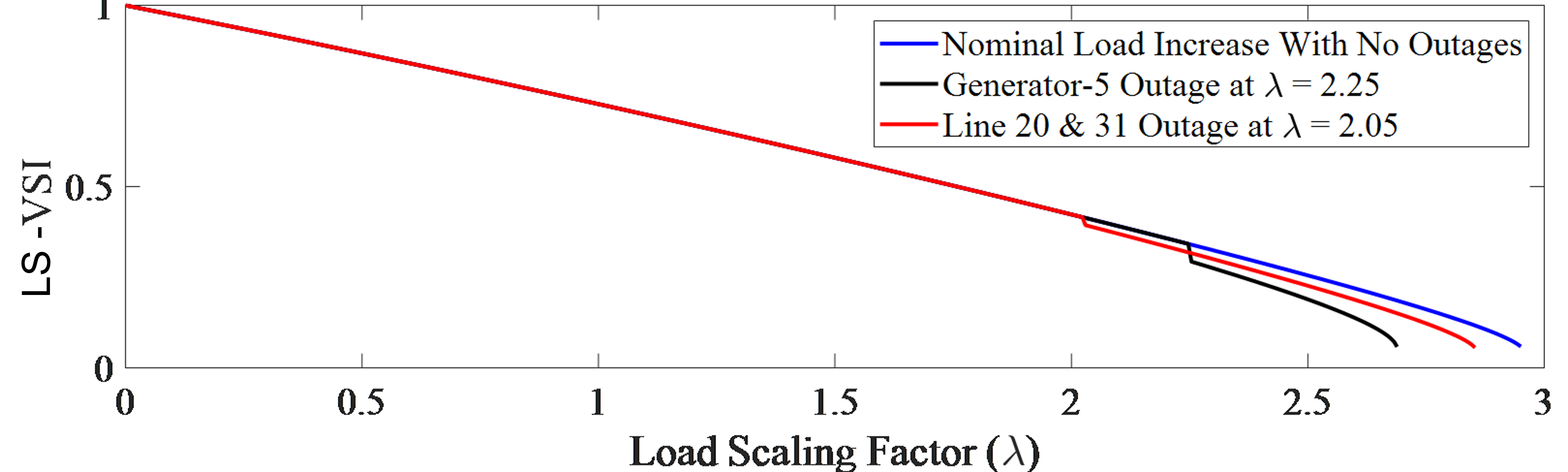}
    \caption{{Impact on LS-VSI at the monitored substation (bus $30$) due to disturbances in unmonitored areas}.} %\textcolor{red}{What do you mean by various system disturbances?}}
    \label{fig:dist_pic}
    \vspace{-2em}
\end{figure}

\subsection{Proposed decentralized LS-VSI versus decentralized and centralized VCPIs}
% The power system operator generally monitors the power system from a control room. 
In practice, there are specific regions of the power system that are critical for LTVI and can be identified by offline means. Thus, the system operator is interested in online LTVI monitoring at a few strategic locations/regions. For instance, in the IEEE-$30$ bus system, the operator would like to monitor the bus $30$ (the critical bus \cite{Guddanti_tsg_ref}). To calculate centralized VCPIs \cite{Wang11} at bus $30$, the voltage measurements at all the buses are required. To calculate the proposed LS-VSI at bus $30$, the voltage measurements at bus $30$ and adjacent branch current measurements are sufficient. %Thus, only one bus type PMU at bus $30$ is required to calculate the LS-VSI.

%Additionally, the proposed Algorithm~\ref{alg:algo} does not require the complete knowledge of {the} system admittance matrix. Instead{,} the proposed Algorithm~\ref{alg:algo} takes advantage of the sparse nature of the power system by using only the admittances of adjacent branches connecting to bus $30$ to calculate the LS-VSI \eqref{eq:final_vsi}.
Tab.~\ref{tab:dist_comp_centr} compares the requirements of centralized VCPIs, a decentralized VCPI (LTI \cite{Vu99}), and the proposed LS-VSI to monitor bus $30$ (the critical bus). \textbf{It can be observed from Tab.~\ref{tab:dist_comp_centr} that the proposed LS-VSI is robust to noise in field measurements (validated in Section~\ref{subsec:noise_simulation_subseciton}), requires only one device (for local measurement), and uses model information about adjacent branch admittances. The decentralized LTI \cite{Vu99} and centralized VCPIs \cite{Wang11} lack at least one of these desirable criteria.}

% It can be observed from Tab.~\ref{tab:dist_comp_centr} that the eventhough the centralized VCPIs \cite{Wang11, Amar18} are robust to the noisy PMU measurements, they have very ambitious requirements such as PMU at every bus in the network (costly), and the full knowledge of system admittance matrix (not desirable). However, decentralized VCPI \cite{Vu99} require only one PMU and no requirement of system admittance matrix but it is highly sensitive to noise in the PMU measurements. Finally, the proposed decentralized VSI combines the benefits of robustness from the centralized \cite{Wang11, Amar18} and minimal infrastructure requirement from decentralized \cite{Vu99} VCPIs 

% \textcolor{blue}{Kishan: I have to make table with all opponents methods and the PMUs needed by them.}
\subsection{Load increase with various system disturbances}%Effect of various disturbances that are not part of monitored currents}
\label{subsec:linetrip_subsection}
In this subsection, we will demonstrate that the proposed LS-VSI can identify the maximum system loading and can also capture the impact of line outages \& generator outages that are not in the immediate neighborhood of the monitored bus. Three loading scenarios are considered
\begin{enumerate}
    \item Nominal load increase where all loads and generators are increased in proportion to their base value.
    \item Load increase with line $20$ \& line $31$ outaged at $\lambda = 2.05$.
    \item Load increase with generator-$5$ outage at $\lambda = 2.25$.
\end{enumerate}

CPF in MATPOWER is used to simulate the system at increasing load levels (using continuation parameter $\lambda$) and the voltage/currents are used to calculate the proposed LS-VSI. Fig.~\ref{fig:dist_pic} shows the plot of LS-VSI at bus $30$ for the above-mentioned three scenarios versus the load scaling factor ($\lambda$) compared to the base case loading condition ($\lambda=1$). As per the derivation in Section~\ref{subsec:decentralized_vsi_derivation}, we expect the LS-VSI to be near $1$ at no-load and monotonically decrease to $0$ at the SNBP. This pattern is indeed observed in all the scenarios validating the proposed method for monitoring LTVI. Further, the outages in the above scenarios (2) and (3) are not in the neighborhood of bus $30$ but their impact \textit{is captured by the bus voltage and branch currents (measurements at bus $30$) which in turn causes a sudden drop in the value of proposed LS-VSI} when the outages occur. Additionally, Fig.~\ref{fig:dist_pic} shows that the drop in the proposed decentralized LS-VSI implies that an event leading to lesser loading capacity has occurred and this fact can be used to trigger mitigation controls locally. % (black line-$\lambda_{SNBP} = 2.6$, red line-$\lambda_{SNBP} = 2.8$, blue line (no event)-$\lambda_{SNBP} = 3$), 
% \begin{figure}
%     \centering
%     \includegraphics[width=\linewidth]{img_folder/system_dist_pic.png}
%     \caption{{Impact on LS-VSI at the monitored substation (bus $30$) due to disturbances e.g: line and gen. outages in unmonitored areas}.} %\textcolor{red}{What do you mean by various system disturbances?}}
%     \label{fig:dist_pic}
%     % \vspace{-2em}
% \end{figure}%\vspace{-1em}

The behavior of the LS-VSI has a similar behavior for other scenarios such as different load increase direction, generator VAR limits, and capacitor switching. \textbf{Thus, the proposed index can monitor the system stability with system disturbances that are not directly in the monitored bus vicinity, making this method applicable for practical systems.}
\begin{table}
	\centering
	\caption{Comparison between centralized, proposed decentralized LS-VSI and other decentralized method to monitor the LTVI of any single substation in IEEE $30$-bus system.
	}
	\renewcommand{\arraystretch}{1.5}
	\begin{tabularx}{\columnwidth}{|>{\centering} p{2.6cm}|>{\centering} p{1cm}|>{\centering} p{1.375cm}|p{2.15cm}|}
		\hline \hline
		Monitoring Method & \# Monitoring devices & Measurement noise robustness & Model info. req. \\ \hline
		Centralized \cite{Wang11}  & $30$ & Robust & Full $Y_{bus}$ \\ \hline
		Decentralized LTI \cite{Vu99} & {$1$} & Not robust & No  information  \\ \hline
		Proposed LS-VSI & {$1$} & {Robust} & {Adjacent branch $Y$} \\ \hline \hline
	\end{tabularx}
	\label{tab:dist_comp_centr}
	\vspace{-2.6em}
\end{table}

\subsection{Impact of noisy measurements on decentralized LS-VSI, Thevenin-based decentralized and centralized VCPIs}%Effect of Noisy PMU Phasor Data on Proposed VSI Versus Other Methods}
\label{subsec:noise_simulation_subseciton}
%\vspace{-1em}
In this subsection, we compared the proposed local LS-VSI with other methods when there is noise in the field measurements. We show that LS-VSI is robust to noisy measurements when compared with local Thevenin index (LTI) \cite{Vu99} and a centralized Thevenin index (CTI) \cite{Wang11}. LTI \cite{Vu99} uses only local measurements to calculate its index while CTI \cite{Wang11} uses full state and full system admittance matrix to calculate its index. 
% \begin{figure}
%     \centering
%     \includegraphics[scale=0.28, trim={2cm 0 2.4cm 2cm}]{img_folder/letter_noisy_figure_v2.eps}
%     \caption{Effect of noisy data on proposed VSI versus other methods. The standard deviation ($\sigma$) of indices with increasing noise levels are: $ LTI - (0.02,0.05,0.2); VSI - (0.001,0.002,0.003); CTI - (0.001,0.002,0.008)$.}
%     \label{fig:noisey_pic}
%     % \vspace{-2.5em}
% \end{figure}

\begin{figure}
    \centering
    % \vspace{-1em}
    \includegraphics[scale=0.36]{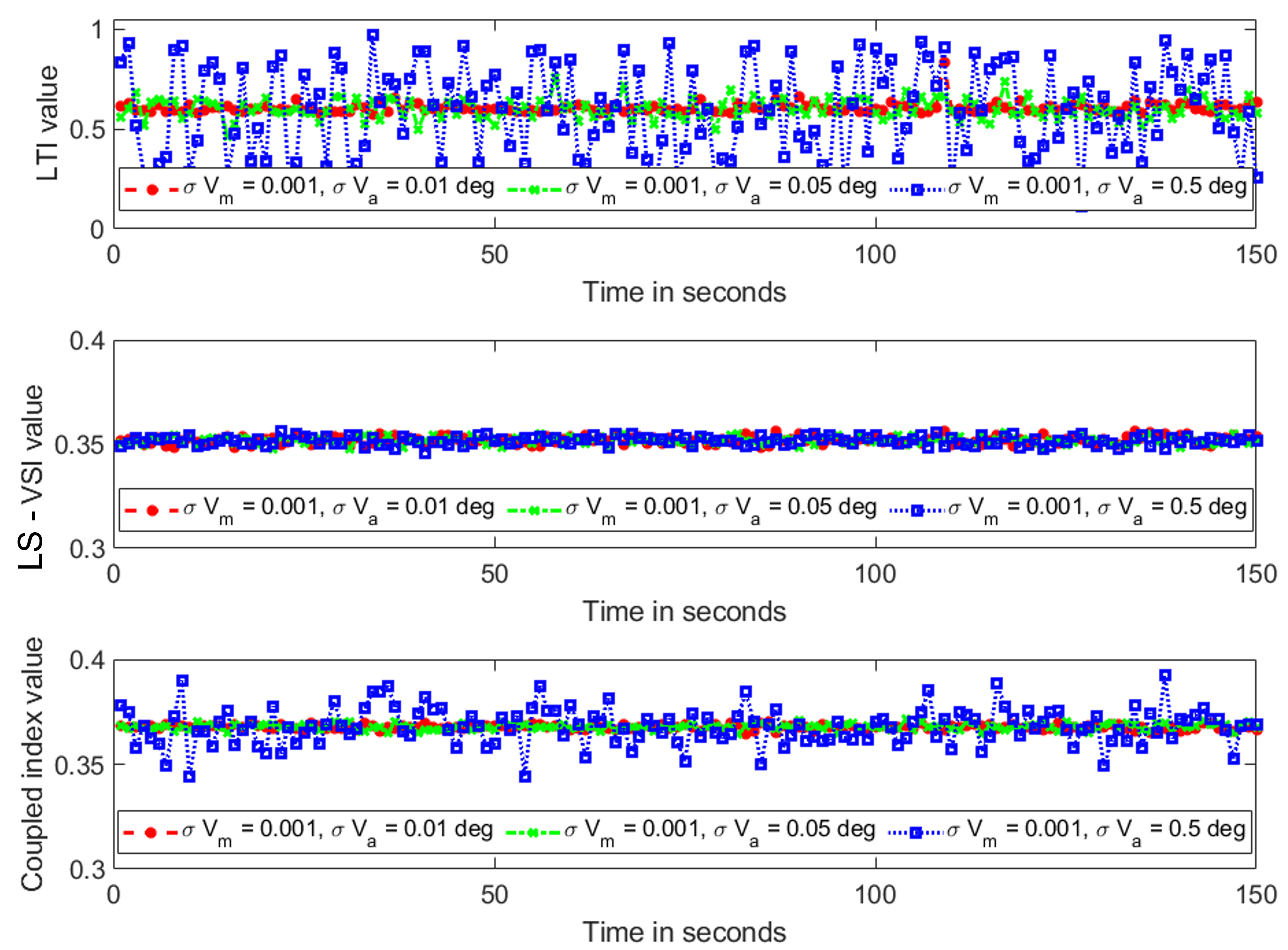}
    \caption{Effect of noisy data on proposed and other methods.}
    \label{fig:noisey_pic}
    \vspace{-1em}
\end{figure}

To understand the impact of noise on the proposed methodology, an additive Gaussian noise with zero mean and standard deviation of $0.001$ p.u. on voltage magnitude ($\sigma V_m$), $0.01^{\circ},\ 0.05^{\circ},\ 0.5^\circ$ on phase angles ($\sigma V_a$) are introduced in the measurements according to the analysis of field-tested PMUs by New England ISO \cite{zhang2013,brown16} and IEEE standard for acceptable PMU errors \cite{martin15}. As shown in Fig.~\ref{fig:noisey_pic}, using the noisy measurements, the LTI, CTI, and the proposed LS-VSI are calculated at bus $30$. Since all indices lie between $0$ and $1$, their variance can be compared and used to quantify the robustness of each method to measurement noise. It can be observed from the plots that the variability of the proposed LS-VSI and CTI is much smaller than LTI. Furthermore, the LS-VSI is {comparable} to that of the CTI, which uses information on the full system state and model information. Tab.~\ref{tab:noise_table_dec_vsi} presents the standard deviation of the indices due to noise in the field measurements. \textbf{Thus, the robustness of the LS-VSI to noisy measurements is better than LTI and is comparable to CTI while needing only local measurements and local system information.} 

%The system is simulated in MATPOWER at a moderate loading level for 150 seconds with random perturbations in the load along a loading direction. 
%Additive Gaussian noise is added to the resulting voltages (V), currents (I), and angles to simulate noisy measurements. The $\sigma$ of noise in the magnitudes of V and I is 0.001 p.u. To showcase how the indices behave with realistic noise, three noise levels in the angles of V and I are considered and they are $\sigma_a$ = ($0.01^{\circ}, 0.05^{\circ}, 0.5^{\circ}$). 
 
\begin{table}[!htbp]
	\centering
	\vspace{-1em}
	\caption{The standard deviation ($\sigma$) for the LS-VSI, LTI and CTI at bus $30$ in IEEE-$30$ bus system when Gaussian noise is introduced in the measurements ($\sigma V_m=0.001 p.u.$).}
	\renewcommand{\arraystretch}{1}
	\begin{tabularx}{\columnwidth}{|P{2.3cm}|P{2.3cm}|P{1.2cm}|P{1.3cm}|}
		\hline \hline
		Noise levels & Proposed LS-VSI & LTI \cite{Vu99} &CTI \cite{Wang11}\\ \hline
		$\sigma V_a=0.01^{\circ}$, \cite{martin15} & $\sigma = 0.001$ & $\sigma = 0.02$ & $\sigma = 0.001$\\ \hline
		$\sigma V_a=0.05^{\circ}$, \cite{zhang2013} & $\sigma = 0.002$ & $\sigma = 0.05$ & $\sigma = 0.002$ \\ \hline
		$\sigma V_a=0.5^{\circ}$, \cite{zhang2013} & $\sigma = 0.003$ & $\sigma = 0.2$ &$\sigma = 0.008$\\ \hline\hline
	\end{tabularx}
	\label{tab:noise_table_dec_vsi}
	\vspace{-1.5em}
\end{table}

\subsection{Identification of critical bus of a system with ZIP loads using proposed decentralized LS-VSI and D-VSI from \cite{Guddanti_tsg_ref}}
\label{subsec:zip_load_simulation}
\begin{figure}
    \centering
    % \vspace{-1em}
    \includegraphics[width=\linewidth]{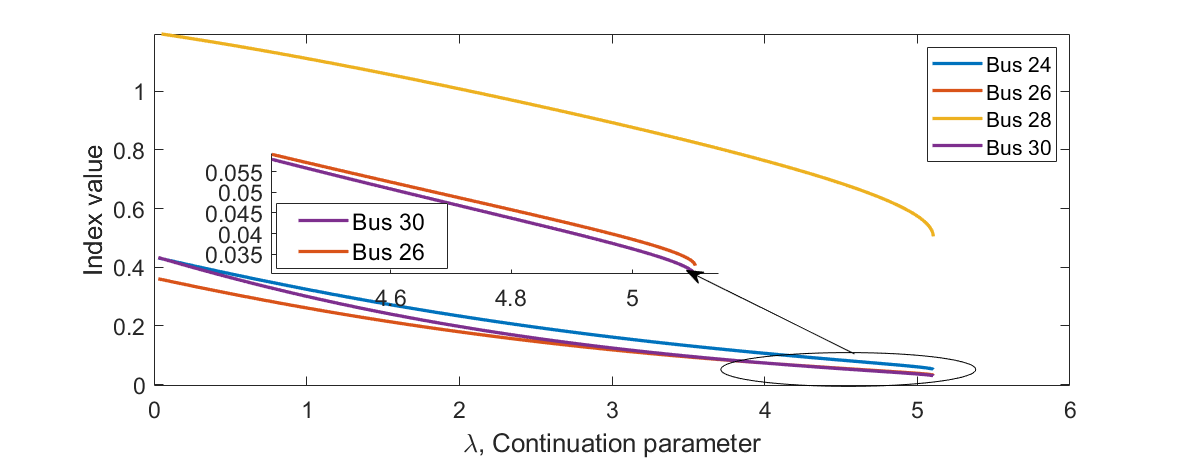}
    \caption{Proposed decentralized LS-VSI at different buses in the network. Bus 30 has smallest value.}
    \label{fig:zip_vsi}
    \vspace{-2em}
\end{figure}

In this subsection, we present the results of the proposed decentralized LS-VSI and distributed index (D-VSI) from \cite{Guddanti_tsg_ref}. The experiment is conducted on IEEE $30$ bus system %\cite{matfile} 
with system wide ZIP loads whose Z, I, P compositions of real and reactive load powers are $\alpha_p=\alpha_q=$ 0.9, $\beta_p=\beta_q=$ 0, $\gamma_p=\gamma_q=$ 0.1 respectively. 

% \begin{figure}
%     \centering
%     \includegraphics[width=\linewidth]{img_folder/pv_curve_zip.png}
%     \caption{PV curve at the critical bus $30$ in IEEE $30$ bus system}
%     \label{fig:pv_curve_zip_1}
% \end{figure}

% Fig.~\ref{fig:pvcurve_lambdap_curve} shows the PV curve at bus $30$, the critical bus that introduces the LTVI in the network. 
The PV curve is traced by using the CPF solver from MATPOWER \cite{ZimmermanEtAl2011}. As detailed in Section~\ref{sec:impact_of_zip_loads}, the SNBP does not occur at the maximum power anymore due to the presence of the ZIP load models in the network. The SNBP occurs when the continuation parameter $\lambda=5.1$. Since we know that bus $30$ is the critical bus that induces LTVI. When $\lambda=5.1$, the index value at bus $30$ must be the least value when compared to index values at other buses in the network indicating that bus $30$ is the critical bus. \textbf{However, \cite{Guddanti_tsg_ref} incorrectly identifies the critical bus to be bus $26$. This is because of its incorrect assumption of system-wide constant power loads. Whereas the proposed LS-VSI correctly identifies the critical bus as bus $30$ as shown in Fig.~\ref{fig:zip_vsi}.}
% \begin{figure}
%     \centering
%     \includegraphics[width=\linewidth]{img_folder/lambdaP_curve.png}
%     \caption{$\lambda$ (continuation parameter i.e., system load scaling parameter) versus real power at bus $30$.}
%     \label{fig:lambdap_curve}
% \end{figure}
% \begin{figure}
%     \centering
%     \includegraphics[width=\linewidth]{img_folder/bus30_pv_plambda_curves.png}
%     \caption{PV curve and real power versus $\lambda$ curve (continuation parameter i.e., load scaling parameter) at bus $30$.}
%     \label{fig:pvcurve_lambdap_curve}
% \end{figure}

% \begin{figure}[h]
%     \centering
%     % \vspace{-2em}
%     \includegraphics[width=\linewidth]{img_folder/p_vsi.png}
%     \caption{D-VSI \cite{Guddanti_tsg_ref} at different buses in the network. Bus 26 has smallest value.}
%     \label{fig:p_vsi}
%     % \vspace{-2em}
% \end{figure} %Both distributed and decentralized VSIs does not become zero (are very close to zero) at SNBP due to numerical issue in the power flow solver as discussed in \cite{Guddanti_tsg_ref} but theoretically, given the real-time measurements at SNBP operation snapshot the proposed VSI becomes zero.

\subsection{Texas synthetic grid: stressing the sinks in Coast area}
\label{subsec:texas_simulation}
\begin{figure}
    \centering
    % \vspace{-1em}
    \includegraphics[scale=0.4]{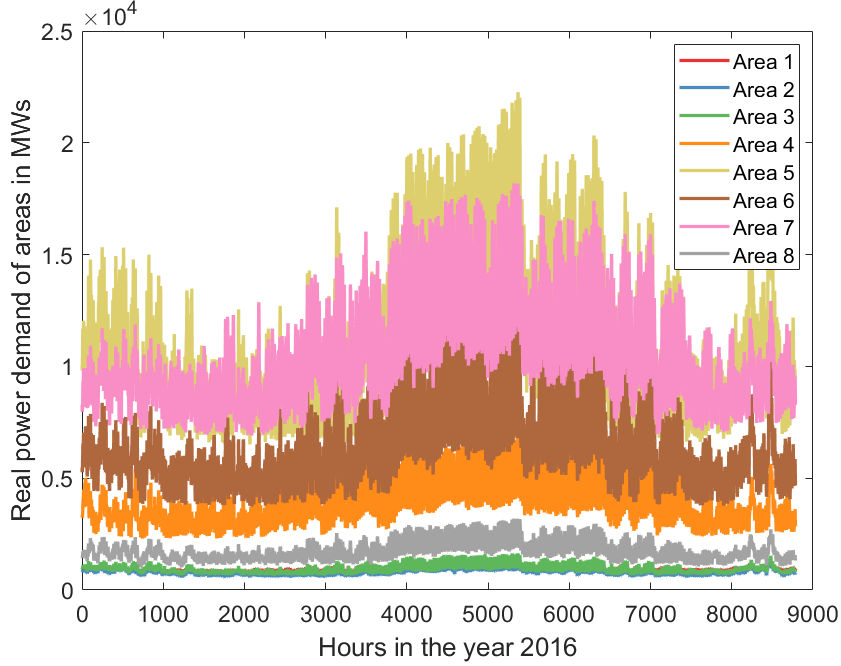}
    \caption{Load profiles of eight different areas in Texas synthetic grid for the year 2016.}
    \label{fig:areas_load_profile}
    \vspace{-2em}
\end{figure}

In this subsection, we present the results of LS-VSI implemented on the Texas synthetic 2000-bus system \cite{Birchfield17, Birchfield172, Gegner16}. The Texas synthetic grid is divided into eight areas with four different voltages levels like 500 kV, 230 kV, 161 kV, and 115 kV. Fig.~\ref{fig:areas_load_profile} presents the load profiles of the eight areas that are designed in a realistic manner for the year 2016 by \cite{Li_tx_18, Li_tx_21}. It can be observed that the area with the greatest demand is area 5 corresponds to the Coast area \cite{Birchfield17}.

\begin{figure}
    \centering
    % \vspace{-1em}
    \includegraphics[scale=0.4]{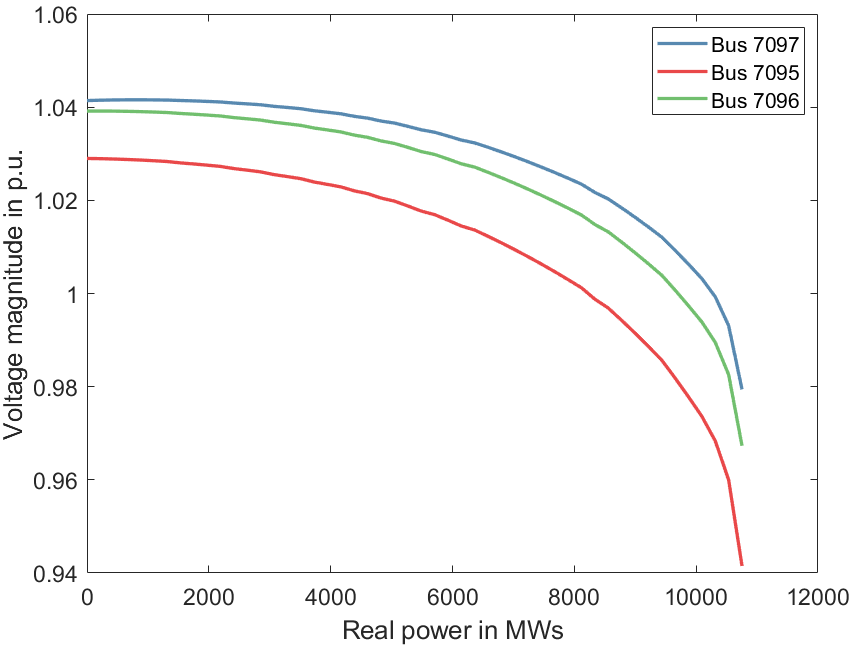}
    \caption{PV curves at different buses when the loads in area 5 are increased linearly.}
    \label{fig:tx_pv_curves}
    \vspace{-1em}
\end{figure}

It is a standard practice to study the voltage stability phenomenon for large interconnected grids by stressing the area of concern and observing the imports and exports between the areas as discussed in \cite{Sapkota09, Randhawa08}. Here, the loads in area 5 are increased linearly using the CPF from MATPOWER \cite{ZimmermanEtAl2011}. Fig.~\ref{fig:tx_pv_curves} shows the PV curves at different buses, and bus 7095 (critical bus) located in area 7 has the lowest voltage magnitude among all buses in the system when voltage collapse occurs.

\begin{figure}
    \centering
    % \vspace{-1em}
    \includegraphics[scale=0.4]{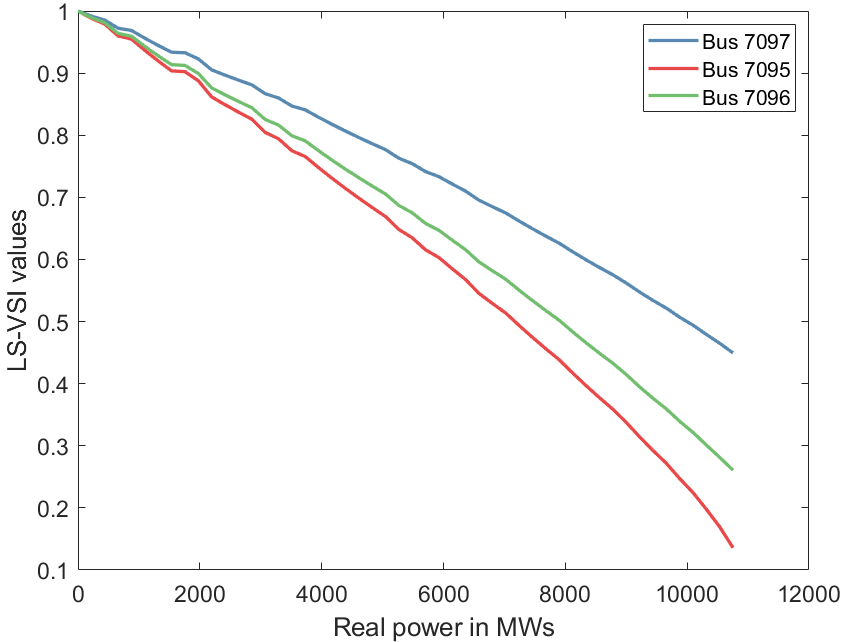}
    \caption{LS-VSI values at different local buses to identify the LTVI phenomenon.}
    \label{fig:tx_lsvsi_results}
    \vspace{-2em}
\end{figure}

It can be observed that the voltage magnitude of the critical bus is 0.94 p.u. when voltage collapse occurs. This is because the practical grids are designed to maintain voltages within a range of 0.9 to 1.1 p.u. for a large set of operating conditions and loading \cite{Birchfield_reactiveplanning}. Therefore, using the voltage magnitude as a signal to trigger local measurement-based SPSs is not always straightforward as it is not a good indicator of system stress. In contrast, the LS-VSI of bus 7095 shown in Fig.~\ref{fig:tx_lsvsi_results} becomes close to zero, indicating a high risk of voltage collapse. \textbf{Thus, LS-VSI values can be computed in real-time by using the local measurements, which are helpful to trigger local measurement-based SPSs reliably.}
	%--
	\section{simulation 2 in PSSE: large disturbance long-term voltage stability}
	\label{sec:sim_2}
\begin{figure}
    \centering
    \vspace{-1.25em}
    \includegraphics[scale=0.5]{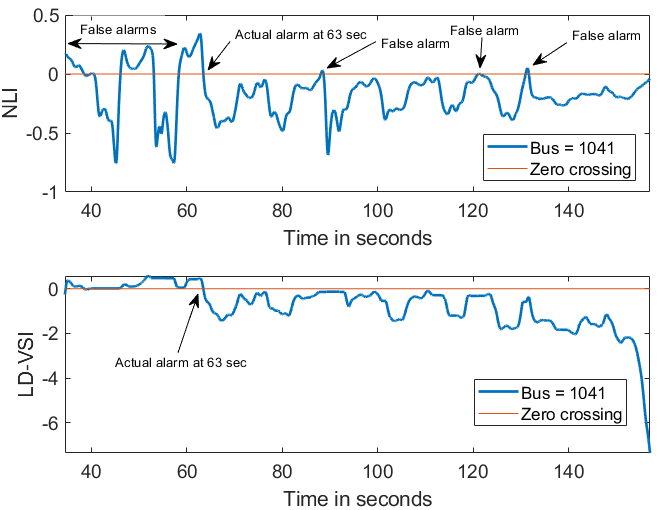}
    \caption{Comparison of NLI \cite{vournas2016voltage} and proposed LD-VSI.}
    \label{fig:nli_vs_our_method}
    \vspace{-2em}
\end{figure}

	In this section, the proposed index (LD-VSI) is validated and compared with NLI \cite{vournas2016voltage} for large disturbance long-term voltage instability in the Nordic system. The IEEE Nordic system is operated in the unstable scenario described in Section~\ref{subsec:nordic_test_system} and simulated in PSSE dynamic simulator. This scenario is explained more in detail (as operating point A) in \cite{taskforce}. To evaluate the impact of measurement noise on LD-VSI, an additive Gaussian noise with zero mean and standard deviation of $0.001$ p.u. on voltage magnitude ($\sigma V_m=0.001 p.u.$), $0.5^\circ$ on phase angles ($\sigma V_a = 0.5^{\circ}$) are introduced according to the analysis of field-tested PMUs by New England ISO \cite{zhang2013,brown16} and IEEE standard for acceptable PMU errors \cite{martin15}. To demonstrate the robustness of the proposed LD-VSI with regard to noise, it is compared with the NLI method \cite{vournas2016voltage} and results are presented in Fig.~\ref{fig:nli_vs_our_method}. Both NLI and LD-VSI methods use the same moving average filter to mitigate the impact of noisy measurements as described in the reference \cite{vournas2016voltage}. The expected behaviors of LD-VSI and NLI are that they both transition from positive to negative during a voltage emergency situation as explained in Section~\ref{subsec:charcateristic_curve} and Fig.~\ref{fig:charac_curves}.

As shown in Fig.~\ref{fig:nli_vs_our_method} both NLI and LD-VSI correctly identify the voltage emergency situation at $63$s but NLI is observed to be noisier. For example, between $40$s and $60$s, NLI triggers false alarms at $40$s and $55$s. However, LD-VSI does not trigger such false alarms before correctly identifying the voltage emergency situation correctly at $63$s. This robustness to noise makes it easier for the operator to reliably implement and identify the voltage emergency situations without having to determine the optimal value for the filter window size (hyperparameter) when compared to the NLI method. Furthermore, after $80$s, NLI again swings between positive and negative values creating false alarms. However, that is not the case with the LD-VSI. Finally, it can be seen that the LD-VSI becomes more negative as the system approaches collapse around $160$s. This can also be used by any mitigation scheme to alleviate system stress. This feature is not observed in the NLI as its value does not vary significantly as the system approaches SNBP. \textbf{Thus, the robustness of the proposed LD-VSI, its smaller filter window size, and its behavior as the system approaches SNBP make it more favorable to use in comparison with similar methods like \cite{vournas2016voltage,vournas2008local}.}

% Additionally, the proposed method has very useful features such as the 1) LD-VSI once it triggers an alarm $63$s by becoming a negative value then it continues to be a negative value making the alarm more obvious to identify, and 2) LD-VSI is theoretically guaranteed to tend to negative infinity at voltage collapse (SNBP) point since its denominator becomes zero at SNBP i.e., $\Pi_1 = 0$ at SNBP as explained in Section~\ref{subsec:decentralized_vsi_derivation}. \textbf{Whereas, NLI does not have such useful interpretation or guarantee at SNBP. Thus, the robustness of the proposed LD-VSI and its requirement of smaller filter window size makes it more favourable to use in comparison with other similar methods.}
	%--
% 	\section{degeneration of power flow circles in special cases}
% 	\label{sec:degenerate_circles}
% 	\input{degenerate_circles.tex}
	%--
% 	%--
% \vspace{-1.2em}
	\section{conclusion}
	\label{sec:conclusion}
	This paper proposes two local measurement based voltage stability indices that can monitor long-term voltage instability occurring over various time scales. The key novelty of the first index (LS-VSI) lies in its decentralized nature, ability to incorporate ZIP loads, and robustness to measurement noise. LS-VSI monitors LTVI by recasting the power flow equations as circles and exploiting the relation between the SNBP and the intersection points. LS-VSI can identify LTVI in scenarios with non-local line trips, generator outages, VAR limits, and shunt capacitor switching. It is shown that the power flow equation-based approach of the proposed decentralized LS-VSI makes it robust to measurement noise which adversely affects the other decentralized techniques such as local Thevenin methods. It is also shown that the LS-VSI is a good indicator to identify voltage collapse and reliably trigger local measurement-based SPSs. LS-VSI is compared with a centralized method and is shown to be comparably robust to measurement noise while requiring lesser measurement infrastructure and system information, making it attractive for practical implementation.

Next, we utilized the properties of LS-VSI to develop the second local index (LD-VSI) for identifying the onset of voltage instability due to large disturbances. We demonstrated the advantage of the proposed LD-VSI over the state-of-the-art method on the Nordic test system through dynamic simulations in PSSE. It is observed that the LD-VSI is more robust to measurement noise and can also distinguish between SNBP and maximum power point, making it a superior method to comparable methods in the literature. 

	\bibliographystyle{IEEEtran}
	\bibliography{Kishan}

\end{document}